\pdfoutput=1
\documentclass[twoside]{article}

\usepackage[sc]{mathpazo} 
\usepackage[T1]{fontenc} 
\usepackage{microtype} 
\usepackage{amssymb} 
\usepackage{amsmath} 
\usepackage{dcolumn} 
\usepackage{wasysym} 
\usepackage{comment} 
\usepackage{graphicx} 
\usepackage{epstopdf} 
\usepackage[english]{babel}
\usepackage[margin=2cm,columnsep=15pt]{geometry} 
\usepackage{multicol} 
\usepackage[hang, small,labelfont=bf,up,textfont=it,up]{caption} 
\usepackage{booktabs} 
\usepackage{float} 
\usepackage{hyperref} 
\usepackage{paralist} 
\usepackage{authblk}
\usepackage{abstract} 
\newcommand{\PicSize}{0.92}

\title{\vspace{-15mm}\fontsize{24pt}{10pt}\selectfont\textbf{Characterization studies of Silicon Photomultipliers and crystals matrices for a novel time of flight PET detector }} 

\author[3]{\small{Etiennette Auffray}}
\author[3]{Faraah Ben Mimoun Bel Hadj} 
\author[1,2]{Daniele Cortinovis\thanks{Corresponding author, daniele.cortinovis@desy.de}} 
\author[3]{Katayoun Doroud}
\author[2]{Erika Garutti}
\author[3]{Paul Lecoq}
\author[4]{Zheng Liu}
\author[4]{Rosana Martinez}
\author[4]{Marco Paganoni}
\author[4]{Marco Pizzichemi}
\author[1]{Alessandro Silenzi}
\author[1,2]{Chen Xu}
\author[1,2]{Milan Zvolsk\'y}
\affil[1]{Deutches Elektronen-Synchrotron (DESY), Notkestrasse 85, 22607 Hamburg, Germany}
\affil[2]{University of Hamburg, Luruper Chaussee 149, 22671 Hamburg, Germany}
\affil[3]{European Organization for Nuclear Research (CERN), 1211, Geneva 23, Switzerland}
\affil[4]{University of Milano-Bicocca, Piazza della Scienza 3, 20126 Milano, Italy}

\date{}

\begin{document}

\maketitle

\begin{abstract}
This paper describes the characterization of crystal matrices and silicon photomultiplier arrays for a novel Positron Emission Tomography (PET) detector, namely the external plate of the EndoTOFPET-US system.
The EndoTOFPET-US collaboration aims to integrate Time-Of-Flight PET with ultrasound endoscopy in a novel
multimodal device, capable to support the development of new
biomarkers for prostate and pancreatic tumors. The detector
consists in two parts: a PET head mounted on an ultrasound probe and
an external PET plate. The challenging goal of 1 mm spatial resolution for
the PET image requires a detector with small crystal size, and therefore high channel density: 4096 LYSO crystals individually
readout by Silicon Photomultipliers (SiPM) make up the external
plate. The quality and properties of these components must be assessed before the
assembly. The dark count rate, gain, breakdown voltage and correlated noise of the SiPMs are measured, while the LYSO crystals are
evaluated in terms of light yield and energy resolution. 
In order to effectively reduce the noise in the PET image, high time resolution for the gamma detection is mandatory.
The Coincidence Time Resolution (CTR) of all the SiPMs assembled with crystals is measured, and results show a value close to the demanding goal of 200
ps FWHM. The light output is evaluated for every channel
for a preliminary detector calibration, showing an average of about
1800 pixels fired on the SiPM for a 511 keV interaction. Finally, the
average energy resolution at 511 keV is about 13 \%, enough for
effective Compton rejection.
\end{abstract}

\bigskip

\begin{multicols}{2} 

\section{Introduction}
\begin{figure}[H]
\begin{center}
\includegraphics[width=\columnwidth]{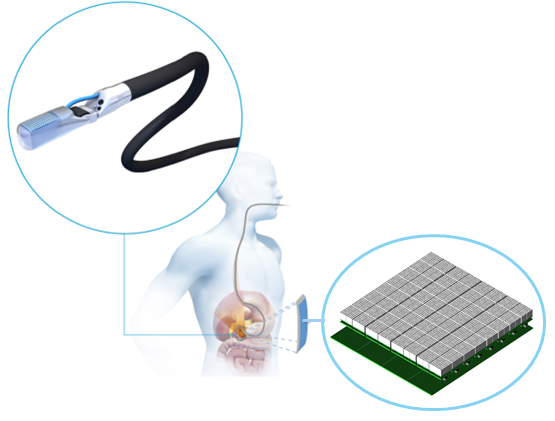}
\end{center}
\caption{The EndoTOFPET-US detector with a magnification of the pancreatic
  endoscope (top circle) and the external plate (bottom circle).}
\label{endotofpet}
\end{figure}

Positron Emission Tomography (PET) is a non invasive, diagnostic
imaging technique for measuring the metabolic activity of cells in the
human body. Nowadays it is widely used for cancer diagnosis
\cite{pet1}. \\
Among the tumors, pancreatic carcinoma is one of the most aggressive
and resistant to current therapies \cite{pancreatic}, and most often is detected only on
an advanced state of development. On the other hand, prostate cancer is
the most common cancer among males \cite{prostate}. \\
Both pancreas and prostate are surrounded by organs with high uptake:
the liver and the heart are close to the pancreas
while the bladder is near the prostate. In these cases the current
full-body multimodal PET scanners have strong limitations, due to the
noise from the neighboring organs and the lack of specific biomarkers
\cite{currentPET} \cite{currentPET2}. 
EndoTOFPET-US \cite{EndoTOFPET} aims to overcome these limits, allowing to investigate the performances of newly developed
specific biomarkers of tumoral processes for pancreas and
prostate. Both organs are usually examined with ultrasound probes through natural cavities, therefore
the diagnostic capability can be enhanced by fusing the morphological
image from ultrasound with a PET metabolic image. The endoscopic
approach and the use of the TOF information with an unprecedented CTR of 200 ps (3 cm along the line of response) allow to
define a specific Region Of Interest (ROI)  and 
significantly suppress  the background from the neighboring organs. \\
The EndoTOFPET-US detector consists of a PET head extension mounted on a
commercial US endoscope placed close to the ROI and an external PET
plate facing the patient's abdomen, in coincide with the PET head
(Figure \ref{endotofpet}).
The external plate is a square of 23 x 23 cm$^2$, and it is made of 256
detector unit modules. Each module consists in a 4x4 LYSO:Ce crystal
matrix glued to a discrete array of 4x4 analog SiPMs from Hamamatsu.\\
LYSO:Ce crystals have high light yield and fast decay time, therefore
appropriate for TOF PET.
SiPMs are novel photon detectors also suitable for TOF PET applications \cite{SiPMpet}, thanks to their compactness, high gain,
insensitivity to magnetic fields, low operating voltage and excellent
timing properties. \\
The following sections describe the experimental setups and the results of the
characterization of the SiPMs, crystals and the combined system
(SiPM + crystal). All the setups have been designed for a fast and
reliable measurement of a large quantity of components, rather than a
detailed study of single devices.


\section{Photodetector Characterization}
The SiPMs chosen for the external plate are matrices of 4x4 discrete
MPPCs (Multi Pixel Photon Counter) from Hamamatsu photonics
(S12643-050CN). Each SiPM in the matrix is a square of 3.0 x 3.0 mm$^2$
with 3464 active pixels (50 x 50 $\mu$m$^2$ each). The distance between two adjacent SIPMs in the matrix (center to center) is 3.6 mm.  These SiPMs exploit the Through Silicon Via (TSV) technology, leading to less dead space and a reduced connection length as compared to conventional wire-bonded SiPM.
The following sections describe the experimental setup, the analysis
method and the results of the characterization of 256 SiPM matrices. 

\subsection{Experimental setup}\label{sec:SiPMsetup}
\begin{figure}[H]
\begin{center}
\includegraphics[width=\columnwidth]{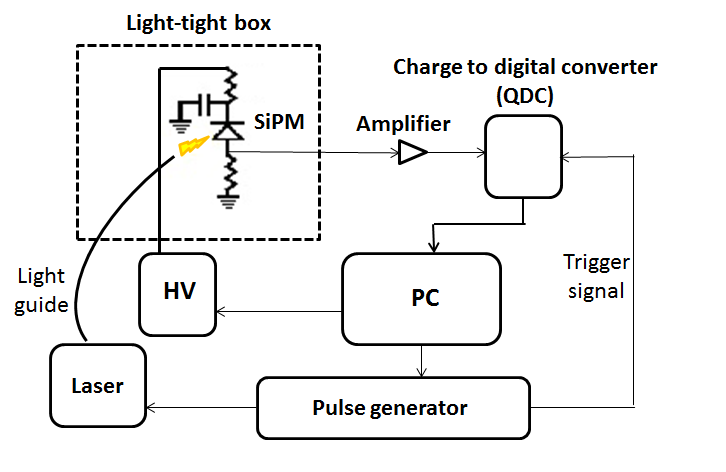}
\end{center}
\caption{Sketch of the experimental setup for the SiPM characterization.}
\label{SiPM_setup}
\end{figure}

The layout of the experimental laboratory setup is shown in Figure \ref{SiPM_setup}. In a light tight box, the SiPM matrix is mounted on a motherboard
incorporating linear amplifiers (Infineon BGA614) and high voltage
filters. The signal for each of the 16 SiPMs of the matrix is obtained using low intensity blue light (Advanced laser diode
system, 451 nm).  The amplified SiPM output is readout by a VME-based
charge-to-digital converter (CAEN QDC965A) with a resolution of 25 fC
per QDC bin. The heat generated by the active components on the SiPM
motherboard is dissipated by a fan. The temperature is monitored by a
sensor (Dallas DS18B20) but not actively controlled. \\

\subsection{Gain and Breakdown voltage}\label{sec:G_Ubd}
\begin{figure}[H]
\begin{center}
\includegraphics[width=\PicSize\columnwidth]{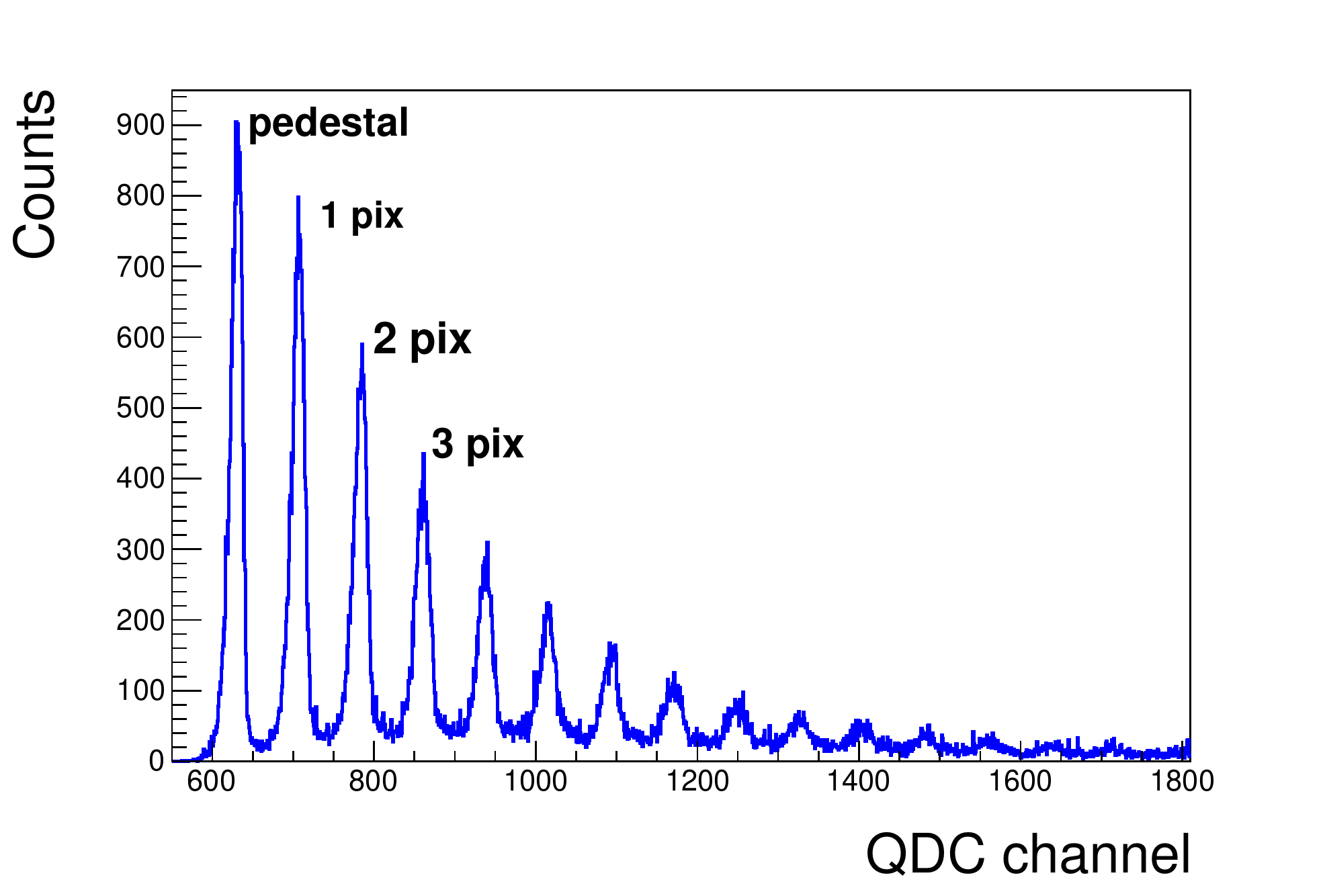}
\caption{Single photo-electron spectrum for a single SiPM, each peak corresponds to a certain number of pixel fired.} 
\label{spes}
\end{center}
\end{figure}

\begin{figure}[H]
\begin{center}
\includegraphics[width=\PicSize\columnwidth]{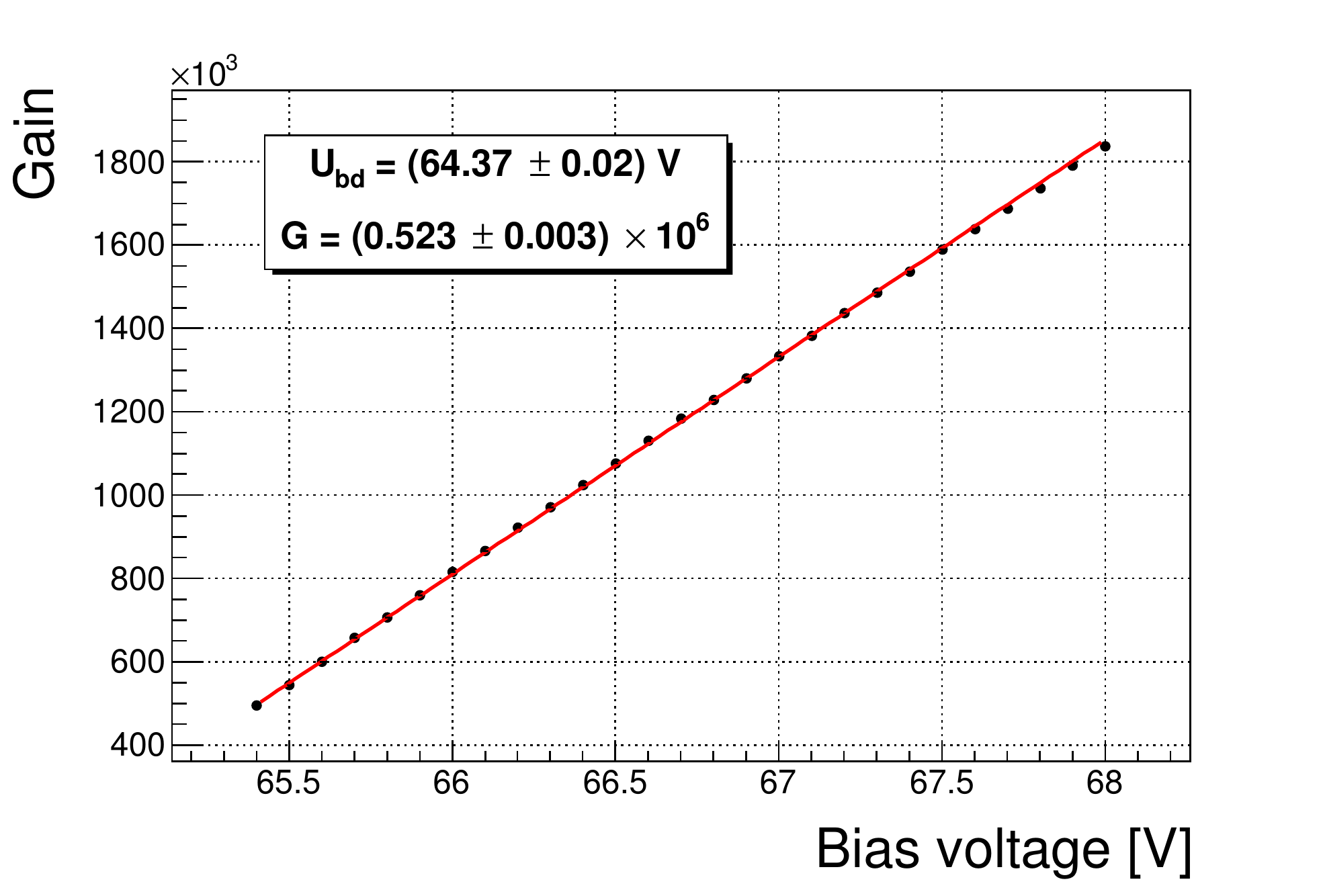}
\caption{Gain as a function of the applied bias voltage for a single SiPM. The red line is
a linear fit on the data points whose errors are within the dots. The
breakdown voltage ($U_{bd}$) and gain ($G$) at 1 V excess bias obtained from the fit are given in the inlet.}
\label{gainfit}
\end{center}
\end{figure}

\begin{figure}[H]
\begin{center}
\includegraphics[width=\PicSize\columnwidth]{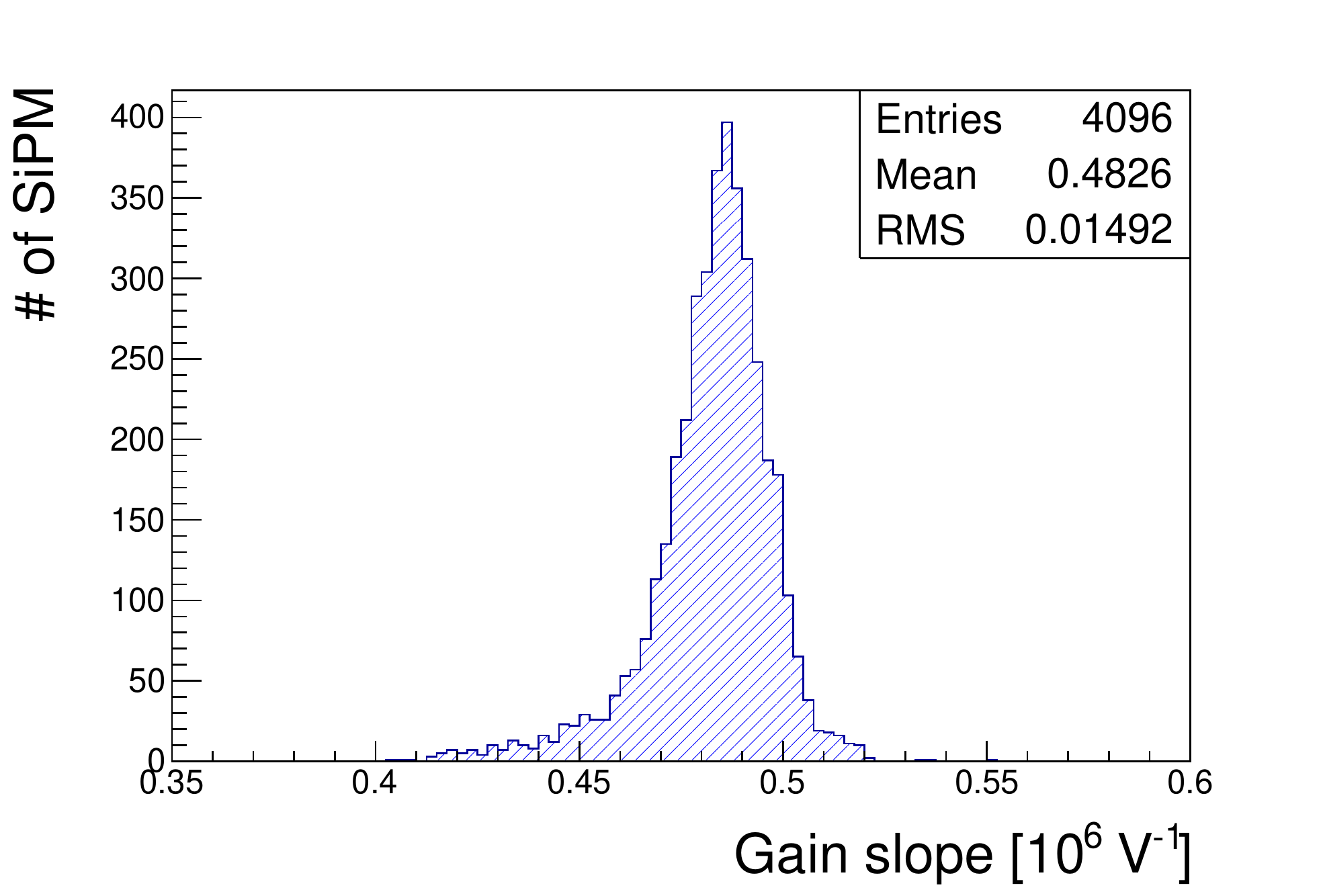}
\caption{Distribution of the gain slope (gain per volt) for all the SiPMs.}
\label{gainHist}
\end{center}
\end{figure}

\begin{figure}[H]
\begin{center}
\includegraphics[width=\PicSize\columnwidth]{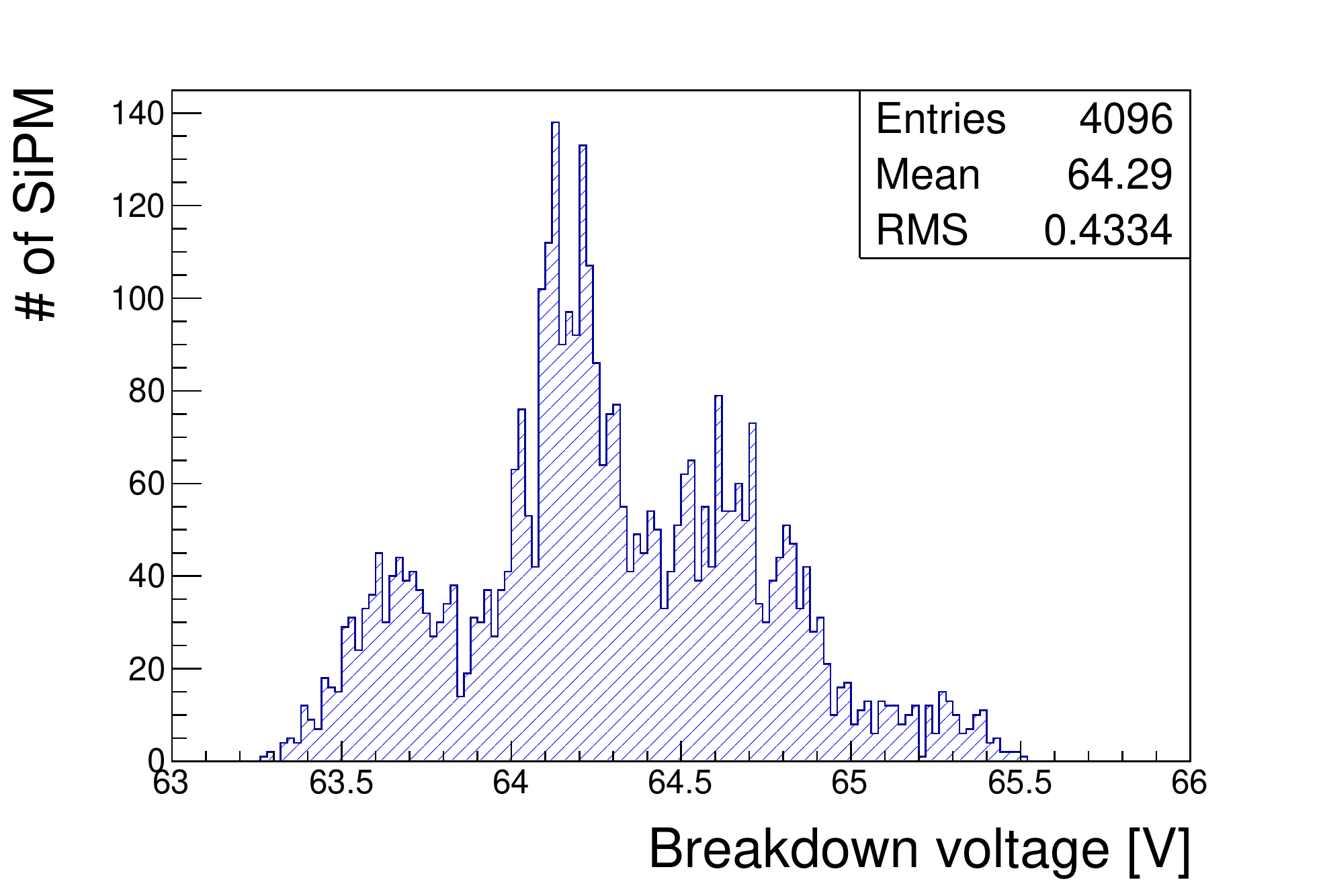}
\caption{Distribution of the extracted breakdown voltage for all the SiPMs.}
\label{UbdHist}
\end{center}
\end{figure}

\begin{figure}[H]
\begin{center}
\includegraphics[width=\PicSize\columnwidth]{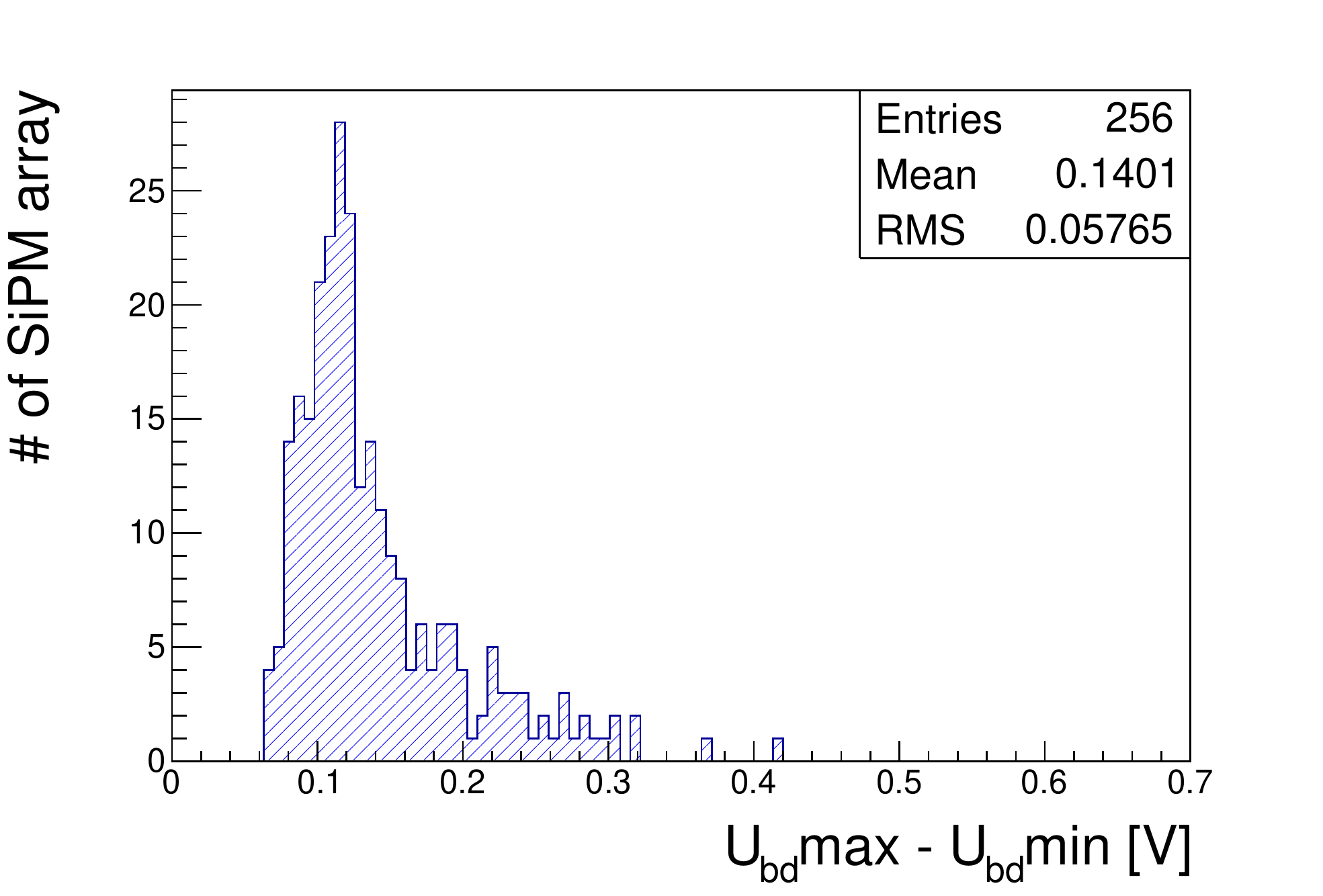}
\caption{Distribution of the maximum $\Delta U_{bd}$, i.e. the difference between maximum and minimum $U_{bd}$ in one matrix.
  The deviation is below the 0.5 V requirement specified to the SiPM producer.}
\label{vbdminmax}
\end{center}
\end{figure}

\begin{figure}[H]
\begin{center}
\includegraphics[width=\PicSize\columnwidth]{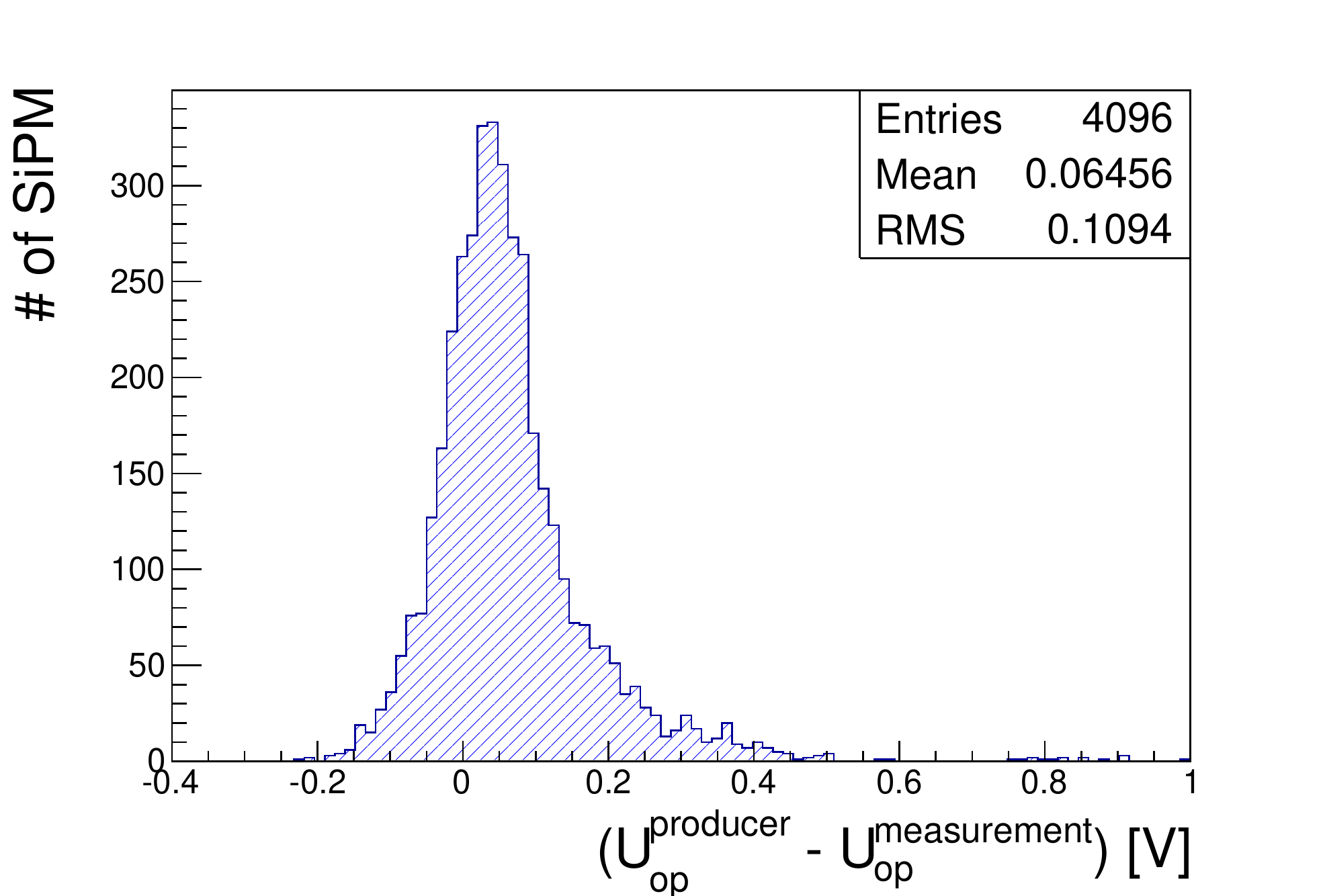}
\caption{Difference between $U_{op}$ given by the SiPM producer and
  $U_{op}$ calculated from the measurement.}
\label{Uop}
\end{center}
\end{figure}

The integrated charge spectrum given by the light pulses in shown in
Figure \ref{spes}. This is repeated for 30 voltage steps, acquiring 200k
events sample per step with an integration gate of 100 ns synchronized
with the laser pulse. The first peak corresponds to the pedestal (zero
pixel fired) while the following peaks correspond to an increasing
number of pixels fired. After the pedestal subtraction, the spectrum is
projected into the frequency space using a Fast Fourier Transform. The
period of the first harmonic is taken as the value of the SiPM gain ($G$) in QDC units, which is then converted into number of electrons by the following formula:
\begin{equation}
 G = \frac{G\cdot R_{qdc}}{A},
\end{equation}
where $R_{qdc}$ is the QDC resolution of 25 fC per QDC bin and $A$ is
the amplification factor of the readout board (19 dB, $\sim$ 8.9).
The error on the gain is about 1\%, obtained as one bin in the Fourier space plus an
additional uncertainty of 5 $\permil$, evaluated by repeating the measurement multiple times under the same conditions. \\
 Figure \ref{gainfit} shows the gain dependence on the voltage bias applied. As expected, the gain is linearly proportional to the bias voltage applied, according to
\begin{equation}
\label{eq:G}
 G =  C_{pixel}\cdot \Delta U = C_{pixel} (U_{bias}-U_{bd} ),
\end{equation}
where $C_{pixel}$ is the pixel capacitance, $\Delta U$ is the excess bias voltage,
$U_{bias}$ is the applied bias voltage and $U_{bd}$ is the breakdown
voltage. Hence $U_{bd}$ is obtained by applying a linear fit and
extrapolating to zero gain, with an uncertainty of the order of a few
tens on mV.
Figure \ref{gainHist} shows the distribution of the gain slope ($G$
at 1 V excess bias) for all the SiPMs. The mean value is
0.48$\cdot$10$^6$ V$^{-1}$ with a spread of 3\% among the SiPMs tested. \\
It is known that the $U_{bd}$ depends on the temperature
\cite{buzhan}, and some fluctuations have occurred during the whole
set of measurements. Moreover, local differences in temperature have
been observed within the SiPM matrix, due to the inhomogeneous
dissipation of the heat generated by the amplifiers. Therefore an
offline temperature correction (see section \ref{tempSection}) has
been applied in order to compare the results.
Figure \ref{UbdHist} shows the distribution of the breakdown voltages
at 25 ${}^{\circ}$C for  all the SiPMs. \\
Although the $U_{bd}$ spread for all the SiPM is about 2 V, this must
not exceed 0.5 V in each single SiPM matrix, because this is the
maximum voltage bias tuning range of the ASIC developed for the SiPM
readout \cite{TOFPET} \cite{STiC}.
As shown in Figure~\ref{vbdminmax}, all the SiPM matrices respect this
requirement. \\
Finally, the results obtained have been compared to the operating
voltages ($U_{op}$) provided by the SiPM producer. The $U_{op}$ is
defined as: $G(U_{OP})=1.25\cdot 10^6$. 
The differences between the $U_{op}$ calculated using equation
\ref{eq:G} with data from the measurements and $U_{op}$ provided by
Hamamatsu are shown in Figure \ref{Uop}. The agreement is good,
despite a shift of 6\% and a spread of 10\%. This is due to the
different measurement method implied by the SiPM producer and the uncertainty of $U_{bd}$ and $G$.

\subsection{Dark Count Rate}\label{secDCR}
\begin{figure}[H]
\begin{center}
\includegraphics[width=\PicSize\columnwidth]{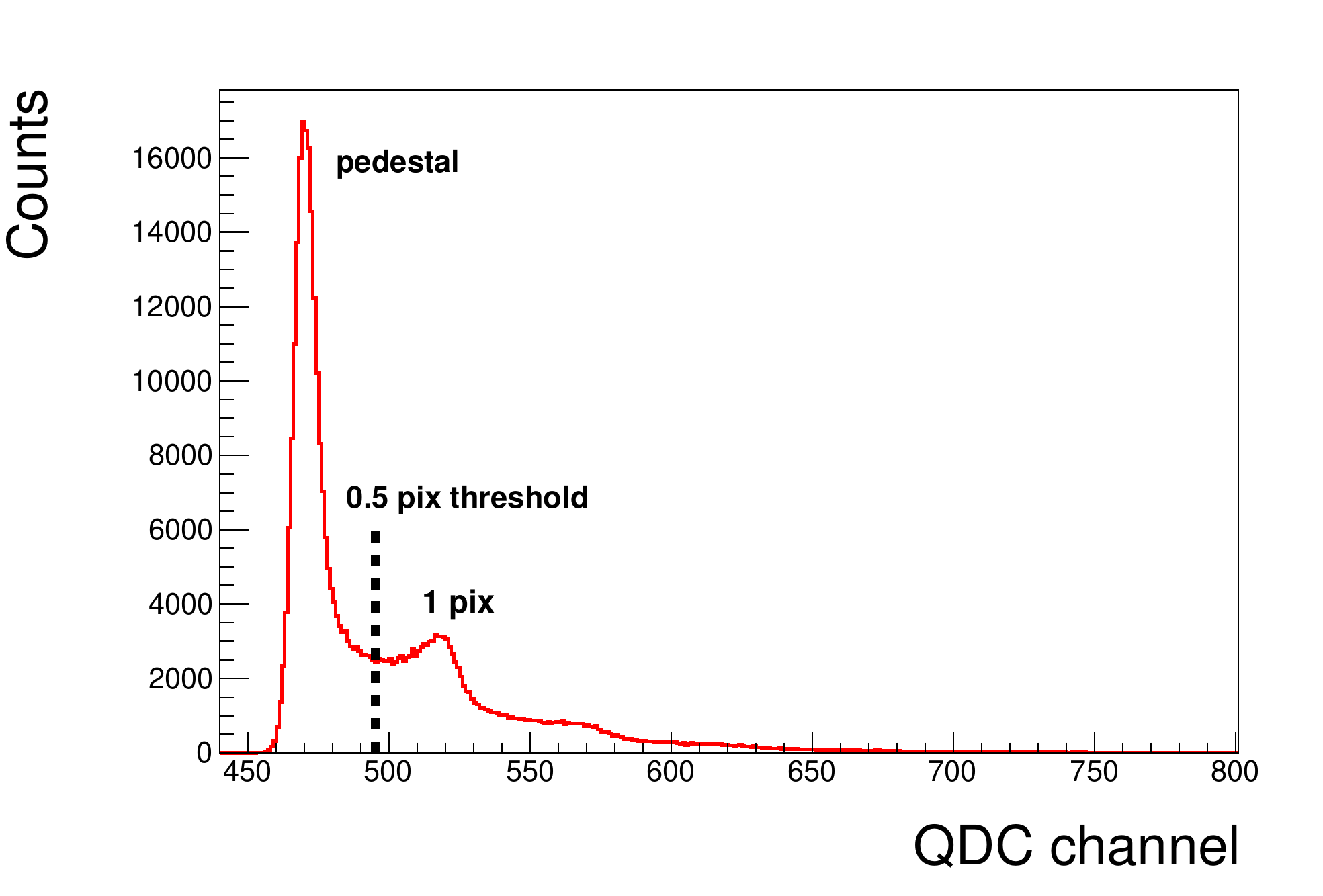}
\caption{DCR charge spectrum. The events below the 0.5 pix threshold
  (pedestal peak) are used to calculate $DCR_{0.5pix}$. }
\label{DCRspes}
\end{center}
\end{figure}

\begin{figure}[H]
\begin{center}
\includegraphics[width=\PicSize\columnwidth]{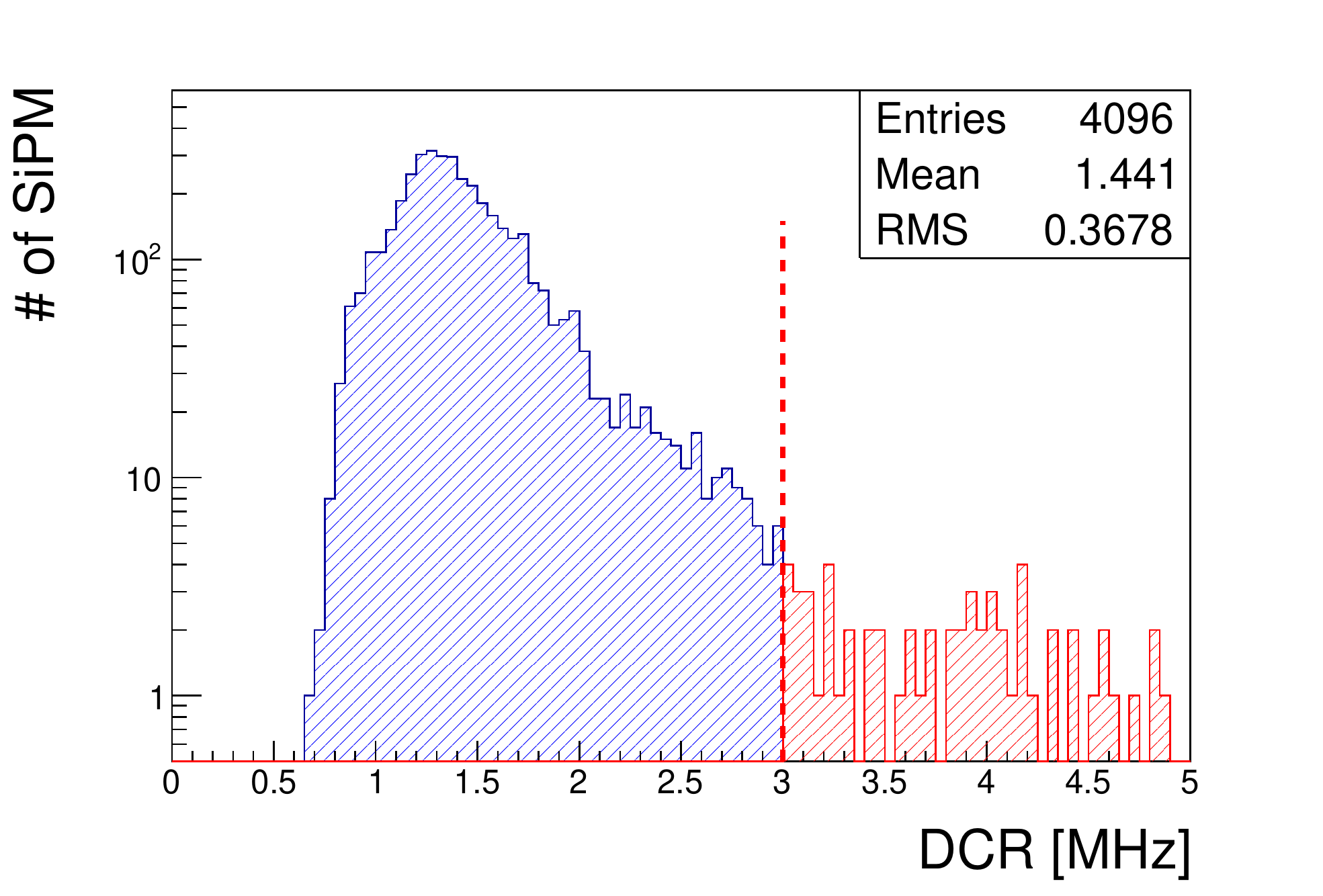}
\caption{Distribution of the $DCR_{0.5pix}$ at 2.5 V excess
  bias and at 25 $^o$C for all the SiPMs. The red line indicates the threshold of acceptable DCR. Entries marked in red correspond to the rejected SiPMs.}
\label{dcr}
\end{center}
\end{figure}

Even in absence of light, an avalanche breakdown can be triggered by a
photoelectron generated from thermal excitation or by tunneling
effect. This phenomenon, usually called Dark Count Rate (DCR), is Poisson distributed and it is measured by randomly integrating the charge with
a time window $\Delta t$ of 100 ns (same time window as for the gain measurement). A typical noise spectrum is visible
in Figure \ref{DCRspes}. \\
The probability of having two dark events in the time $\Delta t$ is
very low, therefore the events corresponding to more than 1 pixel
fired are mainly due to correlated noise, which will be described in
the next section. \\
The events in the pedestal peak, i.e. the events with charge less than
0.5 pixel threshold, are not affected by correlated noise, hence they
are used to calculate the uncorrelated DCR, defined as $DCR_{0.5pix}$. \\
In a time range $ \Delta t$, the expected number of $DCR_{0.5pix}$ events $n$ is given by $n=DCR_{0.5pix}\cdot \Delta t$. The number of events in the
pedestal distribution is given by the Poisson probability density
function for zero $DCR_{0.5pix}$ events in the time $ \Delta t$:
\begin{equation}
P_{0}(\Delta t) = e^{-DCR_{0.5pix}\cdot \Delta t}.
\end{equation}
The $DCR_{0.5pix}$ is therefore calculated using:
\begin{equation}
DCR_{0.5pix} = \frac{\ln(N_{0}/N_{total})}{\Delta t},
\end{equation}
where $N_{0}$ is the number of events in the pedestal peak and
$N_{total}$ is the total number of events recorded.
The error on the calculated value depends on the statistic of the
sample and on the quality of the spectrum. \\
For each SiPM the $DCR_{0.5pix}$ is evaluated for 30 voltage levels, in a range
of 3 V starting about 1 V above the breakdown voltage. 500k
event samples are collected for each voltage point. \\
Also the DCR has a dependence on the temperature, therefore a correction
(see section \ref{tempSection}) has been applied in order to
compensate for temperature variations. \\
The distribution of $DCR_{0.5pix}$, corrected to 25 $^\circ$C and at 2.5 V excess
bias, is shown in Figure \ref{dcr} for all the SiPMs. \\
The DCR has a negative impact on the time resolution. The best time
resolution is achieved when the timing threshold on the readout ASIC
is set as low as possible, up to the noise level.
Therefore a $DCR_{0.5pix}$ acceptance limit
of 3 MHz for each SiPM has been agreed with the SiPM producer. 
Almost all the SiPMs are within the limit, however 5 matrices have been sent back to the producer because they
included at least one SiPM with excessive $DCR_{0.5pix}$.

\subsection{Correlated noise}
\begin{figure}[H]
\begin{center}
\includegraphics[width=\PicSize\columnwidth]{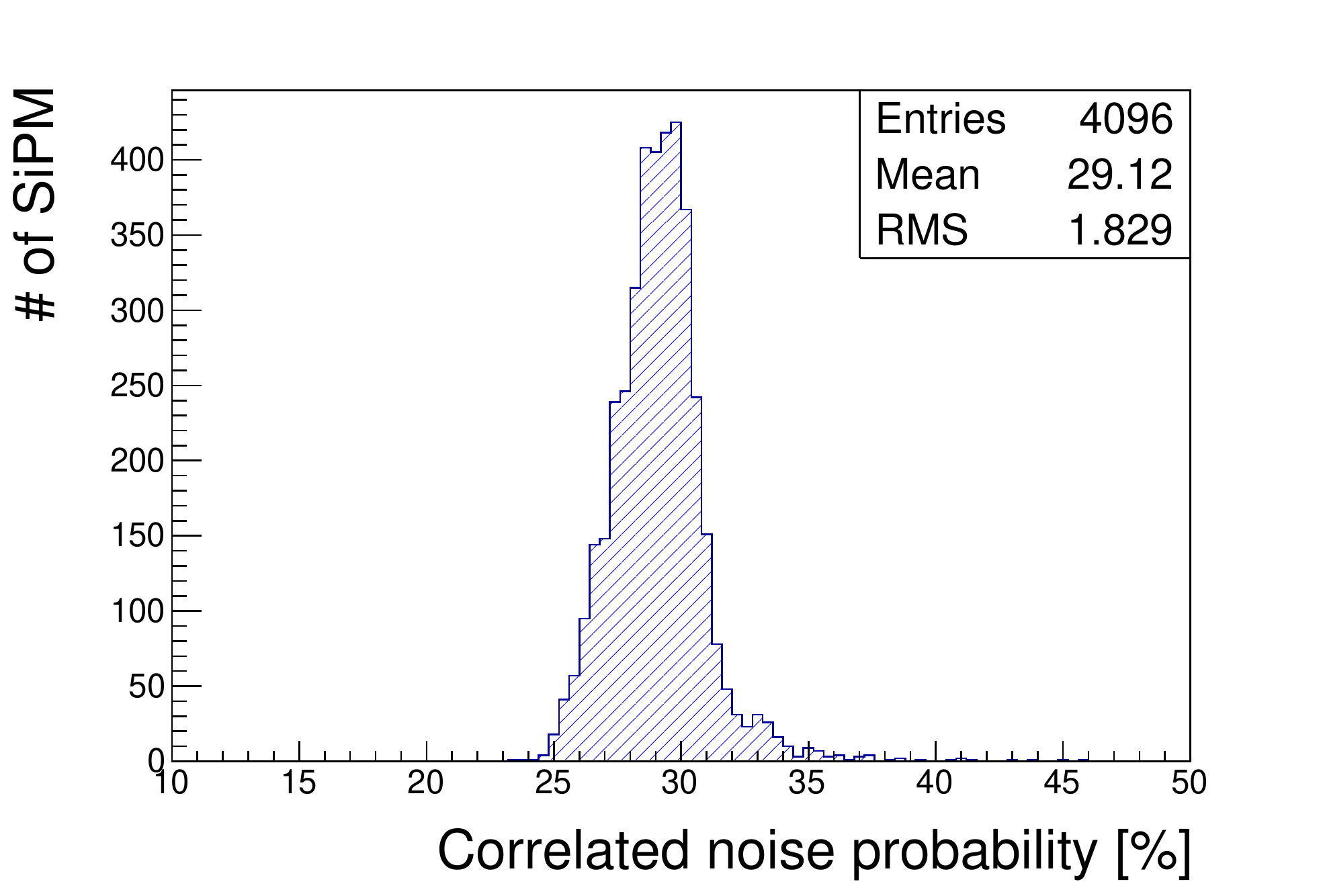}
\caption{Correlated noise probability $P_{cn}$ at 2.5 V excess bias.}
\label{fig:CN}
\end{center}
\end{figure}

Correlated noise includes after-pulse
and inter-pixel optical crosstalk \cite{Hei}. \\
After-pulses are generated if an electron produced during an avalanche
is trapped and released at a later time, with a delay ranging from nanoseconds
up to several microseconds. Depending on the trapping time constant and the pixel
recovery time, trapped electrons can generate an avalanche of equivalent or smaller charge with respect to the DCR pulses. \\
Inter-pixel crosstalk is due to optical photons generated during a
pixel breakdown. These photons have a certain probability to reach the
neighboring pixels, triggering new avalanches. \\
The precise measurement of each of the two effects is beyond the scope of
this study. However, it's possible to estimate the correlated noise
triggering probability $P_{cn}$ as:
\begin{equation}
P_{cn}=\frac{DCR_{1.5pix}}{DCR_{0.5pix}},
\end{equation}
where $DCR_{1.5pix}$ is calculated by setting the threshold to 1.5
pixel fired. Figure \ref{fig:CN} shows the correlated noise
probability at 2.5 V excess bias for all the SiPMs. 
The average results obtained, about 30\%, is compatible with
measurements on similar devices and does not represent an issue for
the detector operation.

\subsection{Temperature dependence}\label{tempSection}
\begin{figure}[H]
\begin{center}
\includegraphics[width=\PicSize\columnwidth]{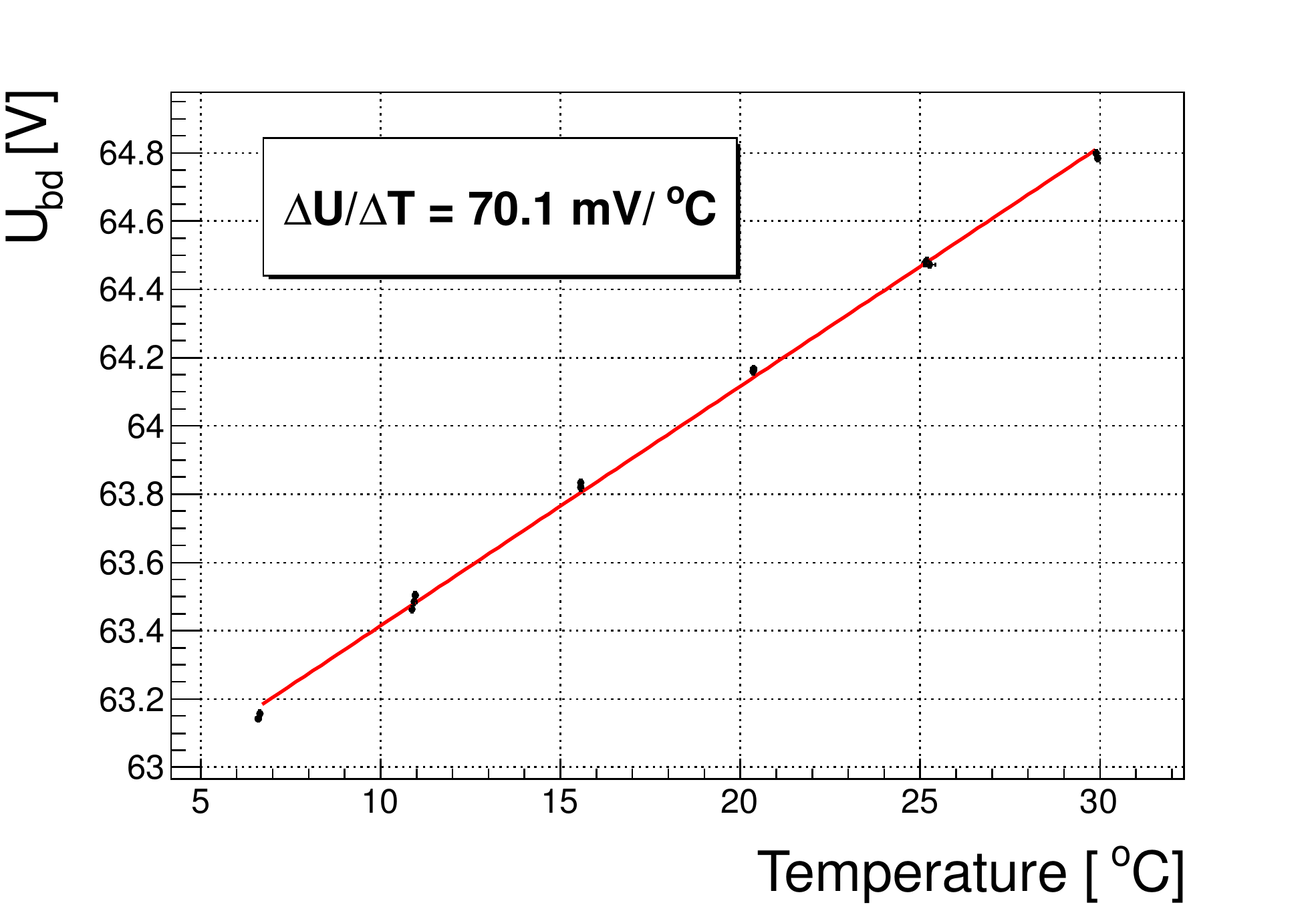}
\caption{Breakdown voltage measured at different temperatures. The
  red curve is the linear fit to the data points. The coefficient
  obtained form the fit is displayed in the inlet.}
\label{vbdvsT}
\end{center}
\end{figure}

\begin{figure}[H]
\begin{center}
\includegraphics[width=\PicSize\columnwidth]{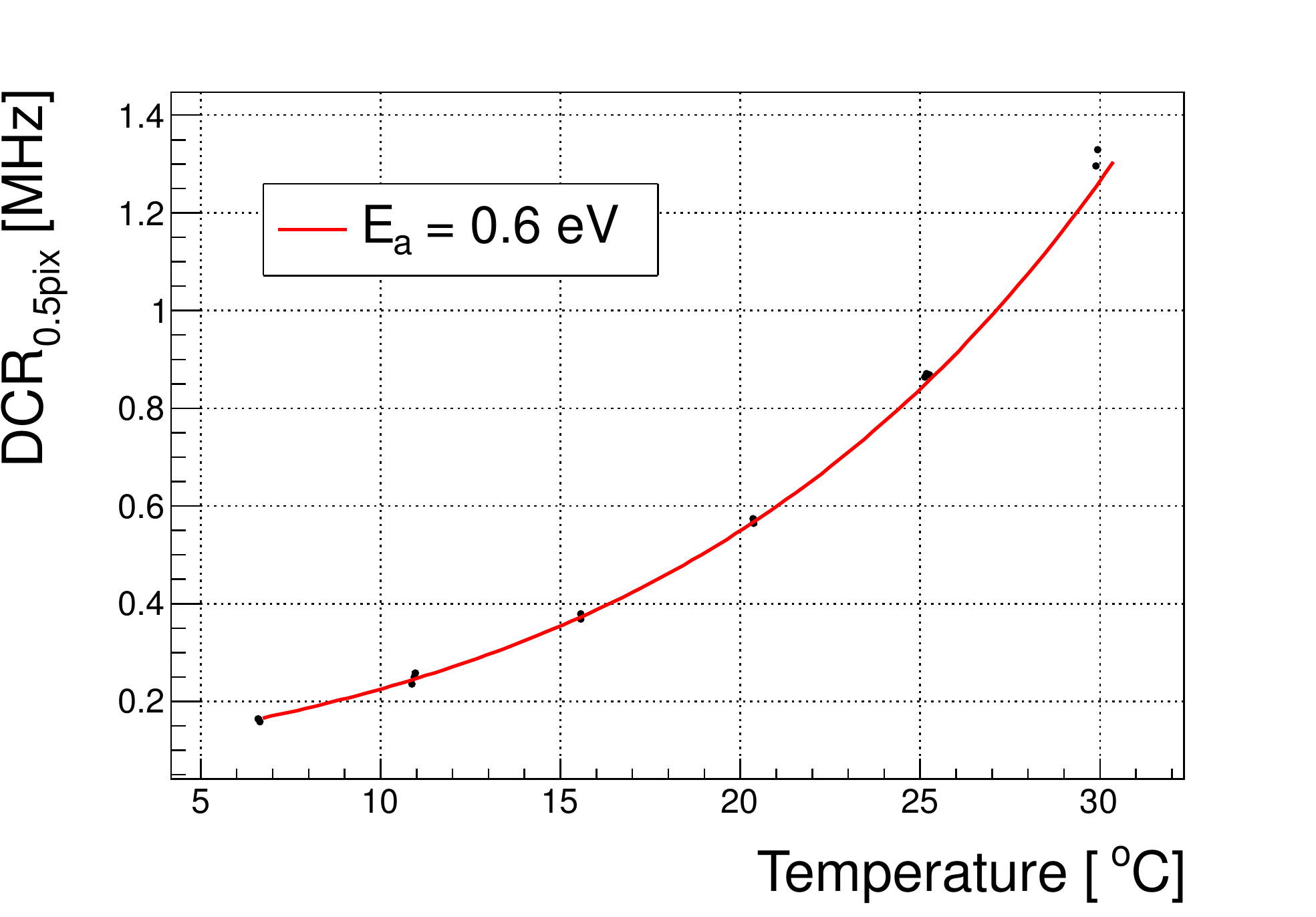}
\caption{$DCR_{0.5pix}$ calculated at different temperatures. The red
line is the fit to data points using equation \ref{eq:boltz}. The
value of $E_a$ obtained from the fit is displayed in the inlet.}
\label{DCRtemp}
\end{center}
\end{figure}

As mentioned in previous sections, SiPM properties depend on
temperature. Therefore, DCR, gain, breakdown voltage and correlated
noise have been measured at different temperatures for a single SiPM in an array.\\   
Using a climate chamber (Espec \mbox{LU-123}), several voltage scans
have been performed in a temperature range between 6 $^{\circ}$C and 30 $^{\circ}$C.
Figure \ref{vbdvsT} shows the linear temperature dependence of the
breakdown voltage; the uncertainty on the temperature is included but
it is smaller than the marker size. The coefficient obtained from the
linear fit is 70.1 mV/$^{\circ}$C, and is used for the temperature corrections applied in
section \ref{sec:G_Ubd} and \ref{sec:LYmodule}. \\
The dependence of the DCR as a function of the temperature should
scale with the density of the thermal carriers \cite{lutz} according to equation:
\begin{equation}
\label{eq:boltz}
n(T)=C_{pixel}T^{3/2}e^{-\frac{E_a}{k_BT}},
\end{equation}
where $C_{pixel}$ is the pixel capacitance, $T$ is the temperature of
the SiPM in Kelvin, $E_a$ is the activation energy and $k_B$
is the Boltzmann constant. \\
Figure \ref{DCRtemp} shows the DCR as a function of temperature at
2.5 V excess bias. An activation energy $E_a$=0.6 eV is obtained by
fitting the data with equation \ref{eq:boltz}. This value is compatible to previous studies \cite{activation}, and it has been used for the temperature corrections applied in section \ref{secDCR}. \\
The change in the gain is found to be less than $1\%/10 ^{\circ}$C,
which is negligible for the temperature variations occurred during the
gain characterization of all the SiPMs. Also the correlated noise
shows a mild dependence on the temperature, but it is not relevant for
the detector performances.

\section{Crystals characterization}

\begin{figure}[H]
\begin{center}
\includegraphics[width=\PicSize\columnwidth]{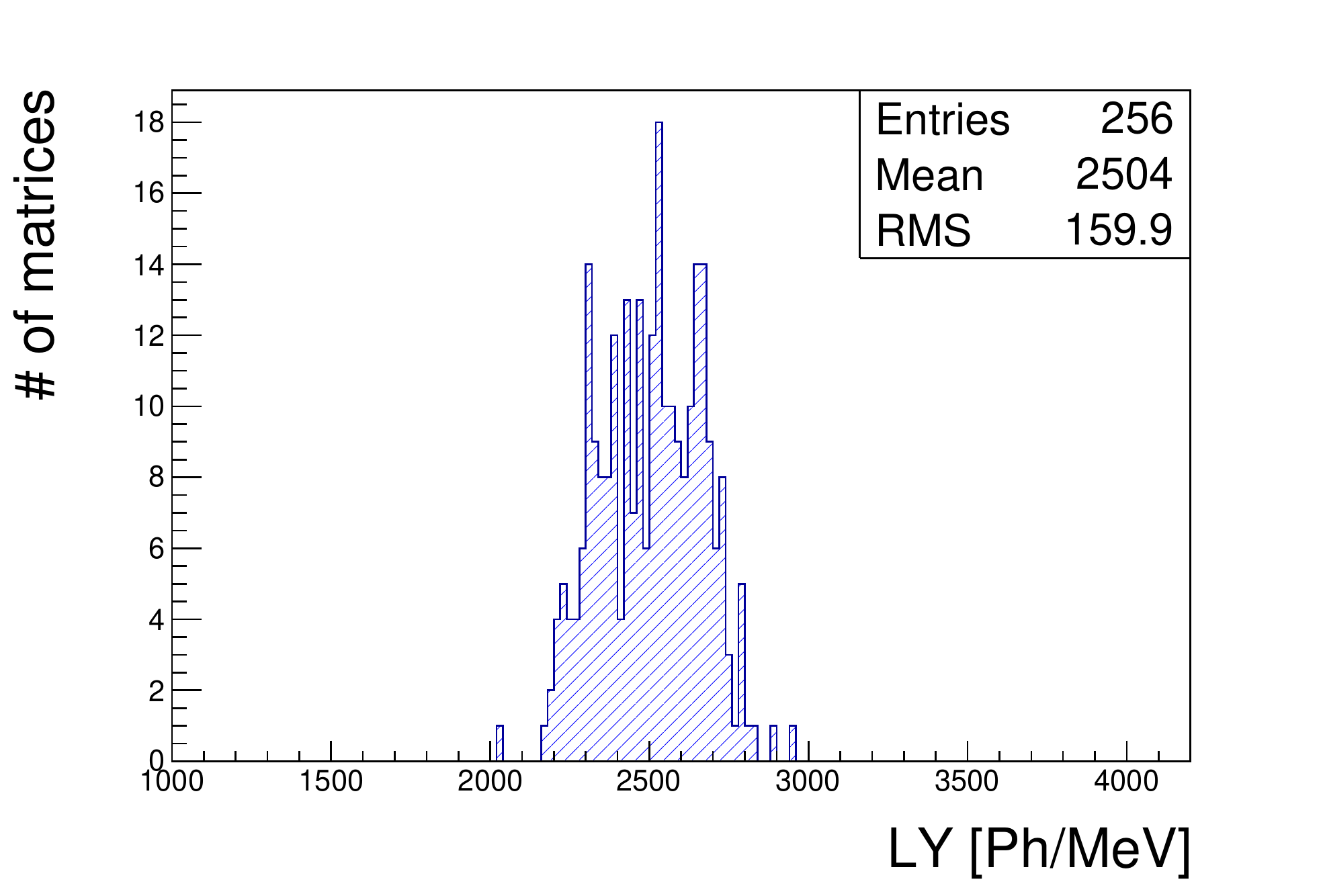}
\caption{Light yield distributions for all the 256 crystal matrices.}
\label{fig:LYminiaccos}
\end{center}
\end{figure}

\begin{figure}[H]
\begin{center}
\includegraphics[width=\PicSize\columnwidth]{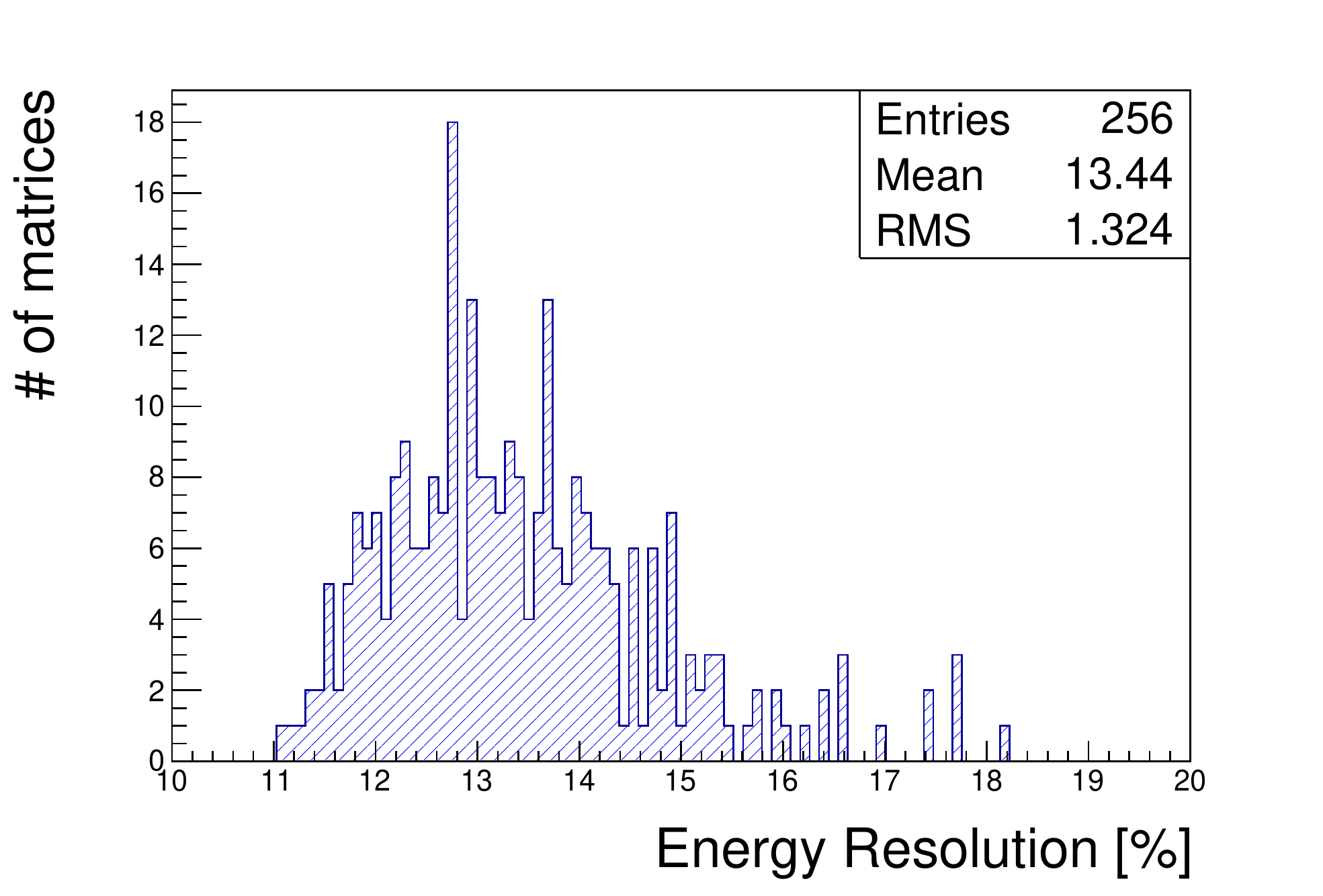}
\caption{Energy resolution distributions for all the 256 crystal matrices.}
\label{fig:ERminiaccos}
\end{center}
\end{figure}

\begin{figure}[H]
\begin{center}
\includegraphics[width=0.74\columnwidth]{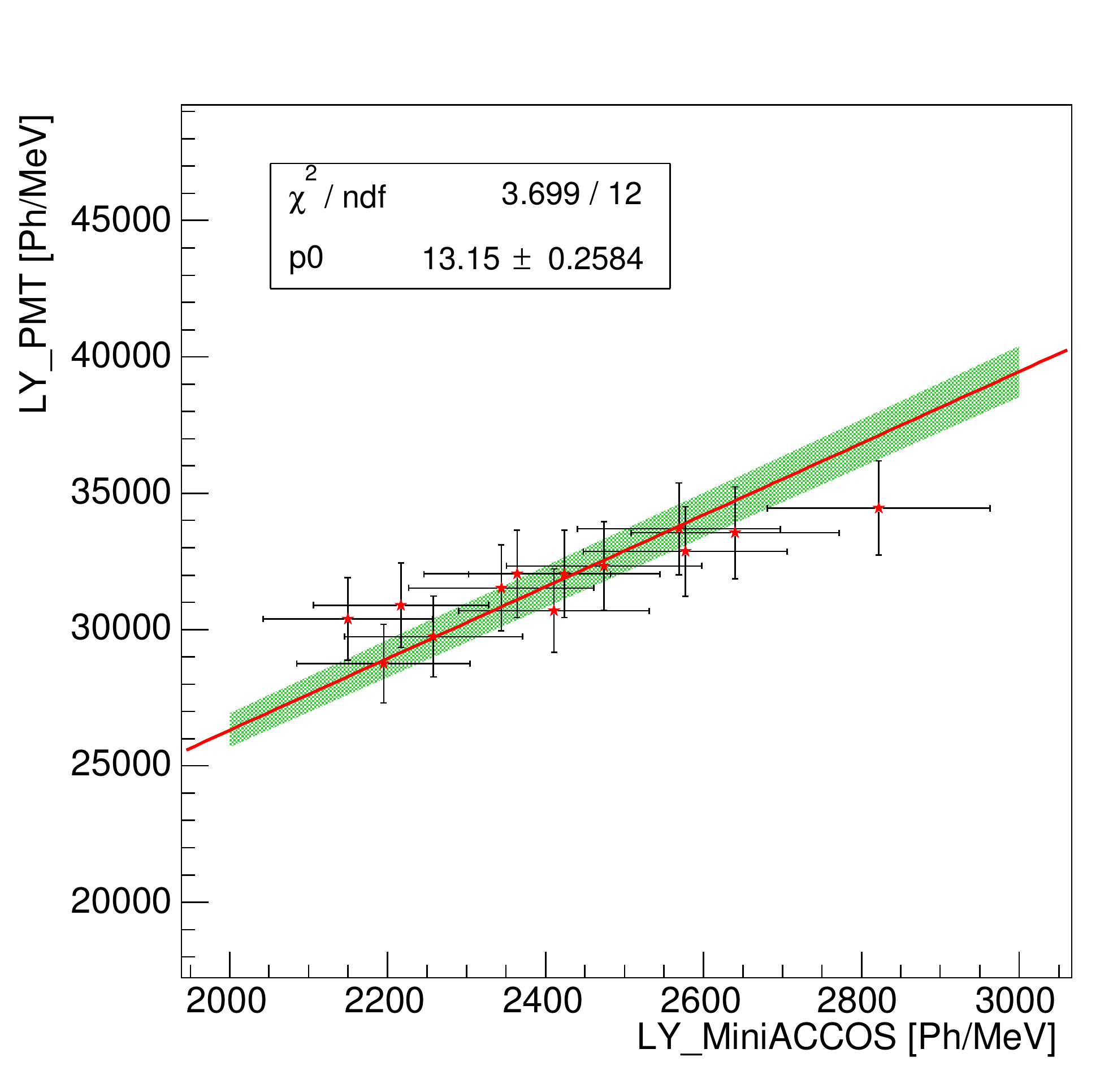}
\caption{Correlated MiniACCOS-PMT data and the lineal fitting function with its 95\% confidence interval band.}
\label{fig:pearson}
\end{center}
\end{figure}

The crystals for the external plate are grouped in 256 matrices of 4x4 LYSO:Ce
scintillators, produced by Crystal Photonics Inc. 
The volume of each crystal is 3.5x3.5x15 mm$^3$ and  within the matrix they are separated by a reflector foil (ESR Vikuiti by 3M). In order to ensure the system uniformity, measurements of light output and
energy resolution have been performed on both faces for each matrix using the MiniACCOS
setup \cite{MiniAccos}, which allows a fast, automatic and highly reproducible data
acquisition process.
In this setup, each matrix is placed on a custom made teflon plate, for a total of 25 matrices
per plate. An air gap of 5 mm is left between the matrix face and the
PMT window. 
Systematic measurements were performed to evaluate uncertainties arising from the bench, yielding a
total relative error of 4.7\% for the light output and 12.8\% for the
energy resolution. 
Figure \ref{fig:LYminiaccos} shows the distribution of light yield obtained with MiniACCOS setup for all the 256 matrices. The distribution spread is 6.4\%,  which subtracting the uncertainty of the setup gives a 4.3\% as
light output dispersion. Figure \ref{fig:ERminiaccos} shows the
distribution of the energy resolution for all the modules, with a mean
value of 13.4\%. The spread is comparable to the bench uncertainty, therefore the variations in
energy resoluton are just due to the setup. \\
In the final configuration, each matrix will be glued to a SiPM array. The absolute values of light
output obtained with the MiniACCOS setup have therefore to be rescaled to this configuration.
In order to obtain this, light output for a subset of 13 matrices was measured on a standard
XP2020Q PMT, using Rhodosil 47V grease as optical coupling and teflon
as back wrapping. The scaling factor between this measurement condition and MiniACCOS was
found to be 13.15, with a correlation coefficient of 0.89 between the two data sets, as shown in
Figure \ref{fig:pearson}. \\
The average light yield for all the crystal matrices can therefore be derived as 32500
$\pm$ 2700 Ph/MeV, including in the evaluation of the uncertainty also the systematics arising from
the calibration process.
Based on the studies reported in \cite{shao}, it can be estimated that at least 20000 $-$ 25000 Ph/MeV are necessary to achieve a CTR
of 200 ps FWHM, which make our set of matrices perfectly suitable for this purpose. The homogeneity on the crystals performance is also ensured within a 4.3\% level for all 256 matrices
tested with an energy resolution of 13.4 $\pm$ 1.3\% as mean value.

\section{Detector Module Characterization}
\begin{figure}[H]
\begin{center}
\includegraphics[width=0.7\columnwidth]{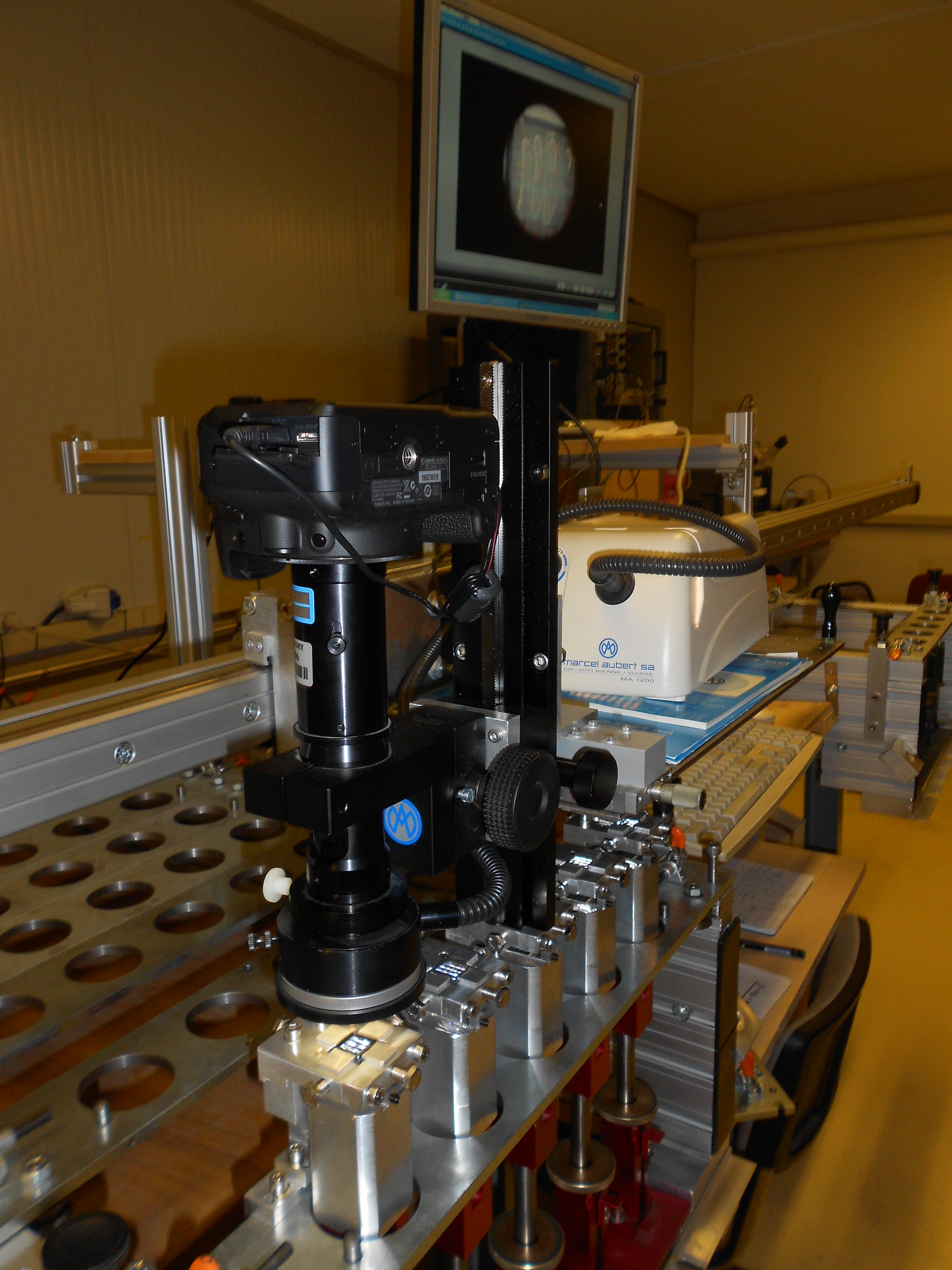}
\caption{The bench for the module assembly. The binocular with the
  camera is placed over the 5 supports for the crystal matrices, the SiPM
  matrices (not visible) are located below.}
\label{fig:glueSetup}
\end{center}
\end{figure}

A dedicated bench has been developed to glue the crystal matrices on
the SiPM arrays, allowing to assemble 5 modules per day.
Figure \ref{fig:glueSetup} shows the setup: the crystal matrices are placed on a
movable support and a camera allows the visual alignment with the
SiPM matrices located under the crystals. The amount of glue (RTV3145)
is calculated in order to have a maximum thickness of 0.1 mm, and it is
spread over the SiPM matrix with a dispenser. The SiPM matrix is then
pushed againts the crystals and a visual inspection is performed to check
if any air bubble occurred. If none, the modules are left on the support for the curing time (24 h). \\
Every module is then measured regarding light output, energy
resolution and CTR, which are described in the next sections.

\subsection{Detector light output}\label{sec:LYmodule}
The detector light output for the 511 keV photon (defined here as $LO_{511}$) is of primary importance for the detector time
resolution. As demonstrated in other studies \cite{shao}, the number
of detected photoelectrons must be maximized in order to achieve the
best time resolution. Moreover, all the channels should have the same
response for the 511 keV photon, allowing an easier detector
calibration. \\
A $^{22}Na$ source with an activity of about 1 MBq is used as a $\beta ^+$
emitter. The setup is similar to the one described in section
\ref{sec:SiPMsetup} but the SiPM signal is not amplified and the
resolution of the QDC is set to 200 fC. The charge integration gate is set to 450 ns and it is triggered by the SiPM signal
itself, using a discriminator (CAEN model 96) and a gate generator
(LeCroy model 222). A channel switcher (Keithley 7002 switch system) provides automatic scan through the 16
channels in each matrix. The temperature is recorded before each
channel measurement. Each SiPM is operated at $U_{op}$ as defined in section \ref{sec:G_Ubd}, eventually corrected for temperature variations
according to the coefficient of 70.1 mV/${}^{\circ}$C (see section \ref{tempSection}).

\begin{figure}[H]
\begin{center}
\includegraphics[width=\PicSize\columnwidth]{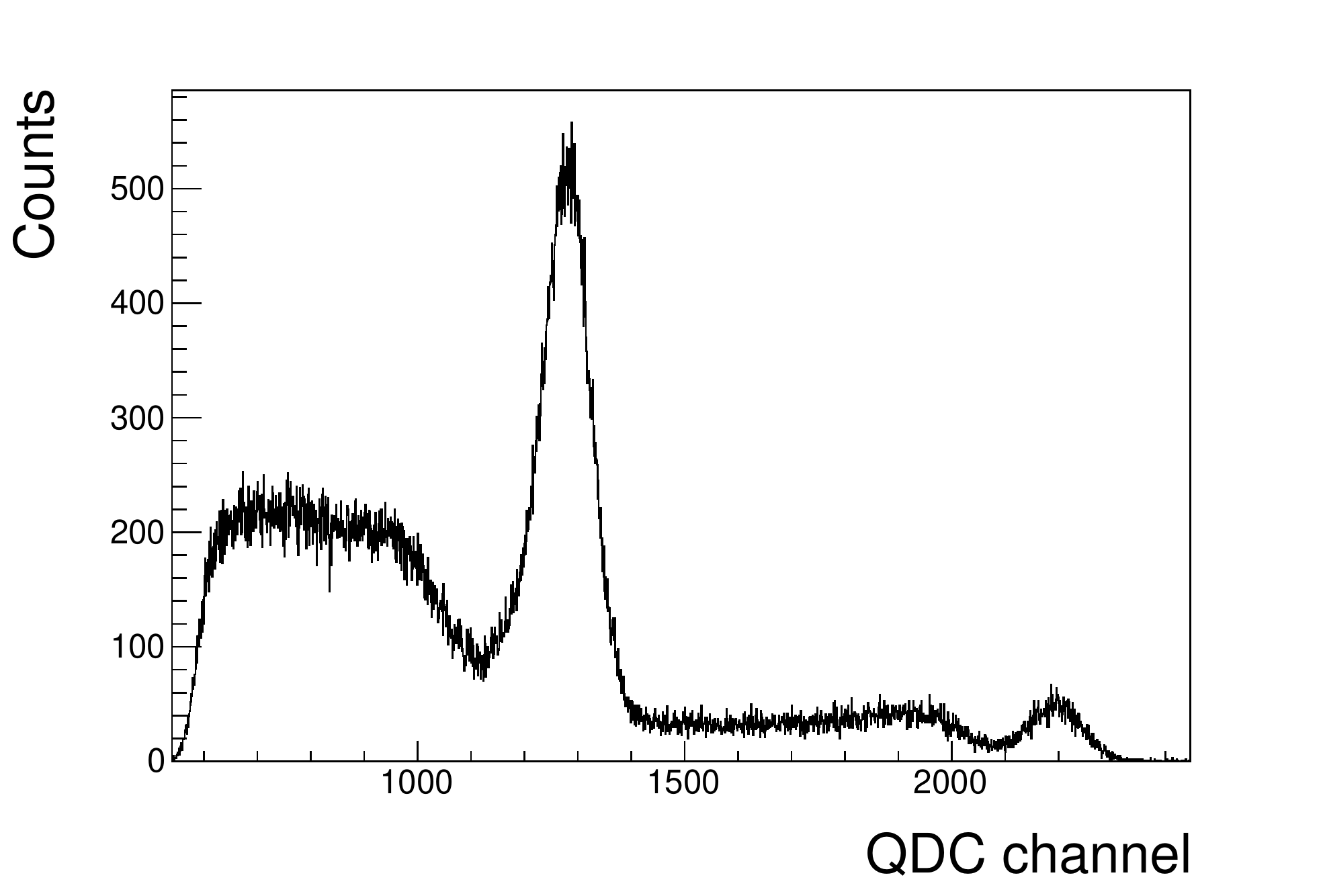}
\caption{$^{22}Na$ spectrum for one channel. The higher peak corresponds to the full absorption of the 511 keV photons, while the peak on the right comes from the 1277 keV gamma emitted during the de-excitation of Neon to the ground state.}
\label{Naspectrum}
\end{center}
\end{figure}
\begin{figure}[H]
\begin{center}
\includegraphics[width=\PicSize\columnwidth]{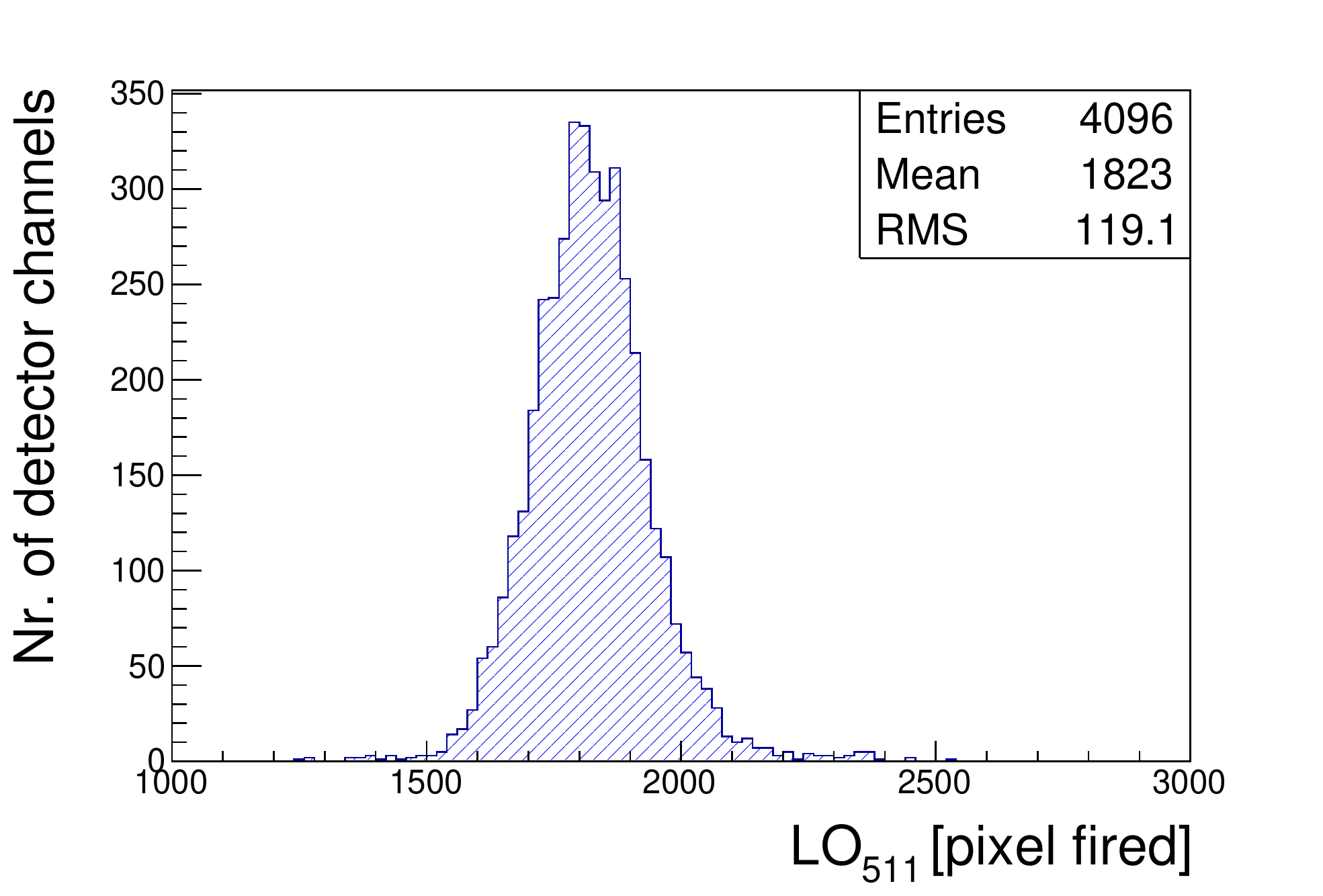}
\caption{Light output distribution for all the channels.}
\label{LOmod}
\end{center}
\end{figure}

Fig \ref{Naspectrum} shows a typical $^{22}Na$ spectrum for a single channel. The peak corresponding to the 511 keV photon in the
charge spectrum is fitted by a Gaussian function. Hence the light
output $LO_{511}$, expressed in number of pixels fired on the SiPM, is obtained by the following relation:
\begin{equation}
\label{eq:LO}
LO_{511}=\frac{(Q-P)\cdot r_{qdc}}{G},
\end{equation}
where $Q$ is the mean value of the Gaussian fit, $r_{qdc}$ is 200 fC
per bin, $P$ is the pedestal position, and $G$ is 1.25x$10^6$.
The uncertainty on $LO_{511}$, of the order of about 30 pixels, is
given by statistical error and temperature uncertainty of 0.5 $^o$C. No error for
the $U_{op}$ is available. \\
The light output distribution for all the channels is shown in
Figure \ref{LOmod}, the average value of about 1800 pixels should
allow to reach the CTR of 200 ps FWHM. \\
A preliminary detector calibration will be based on these measurements.

\subsection{Inter channel cross-talk}\label{sec:Xtalk}
\begin{figure}[H]
\begin{center}
\includegraphics[width=0.50\columnwidth]{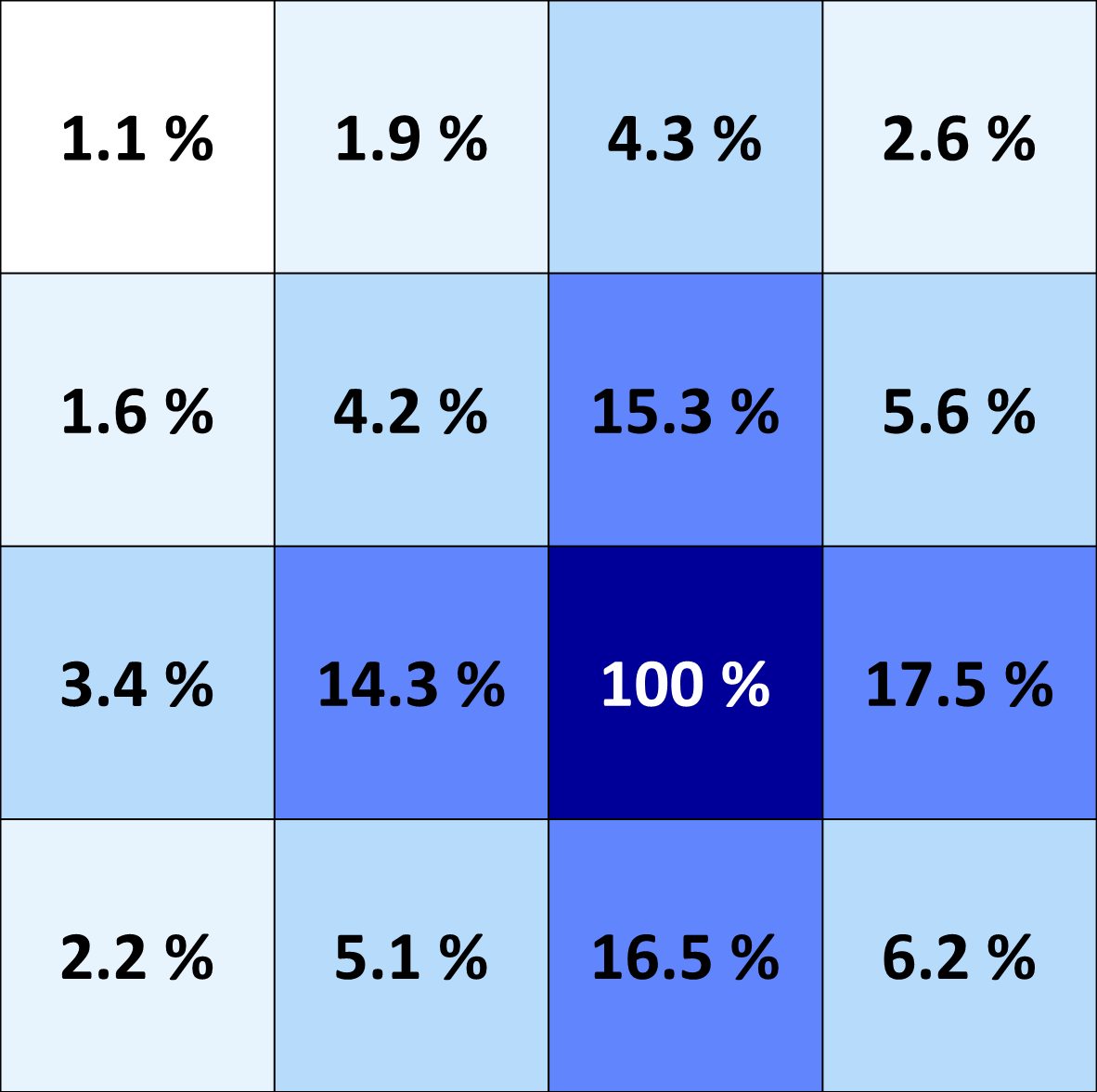}
\caption{Average cross-talk for the 511 keV photon obtained over 4 measurements, using each time a central channel as trigger. }
\label{fig:Xtalk}
\end{center}
\end{figure}

The cross-talk between the module channels is measured on a sample module.
Similarly to the setup described in section \ref{sec:LYmodule}, the $^{22}Na$ source is placed in front of the module and only one channel is chosen as trigger. Any other channel in the module is acquired simultaneously to the trigger, using a common integration gate of 450 ns in the QDC. The channel switcher provides the automatic scan trough the 15 channels, each operated at its own $U_{op}$.\\
In every channel, only the events corresponding to the photopeak in the trigger are selected. Therefore, dividing the mean charge of these events by the photopeak position in the trigger, the cross-talk for the 511 keV photon is obtained for all the remaining 15 channels in the module.
The measurement is repeated by selecting different trigger channels in the module. Figure \ref{fig:Xtalk} shows the average crosstalk obtained from 4 different measurements: every time one of the central channels in the module is used as a trigger. The cross-talk does not exceed 20\%, and can be suppressed by setting a threshold at 100 keV.

\subsection{Energy calibration}
\begin{figure}[H]
\begin{center}
\includegraphics[width=\PicSize\columnwidth]{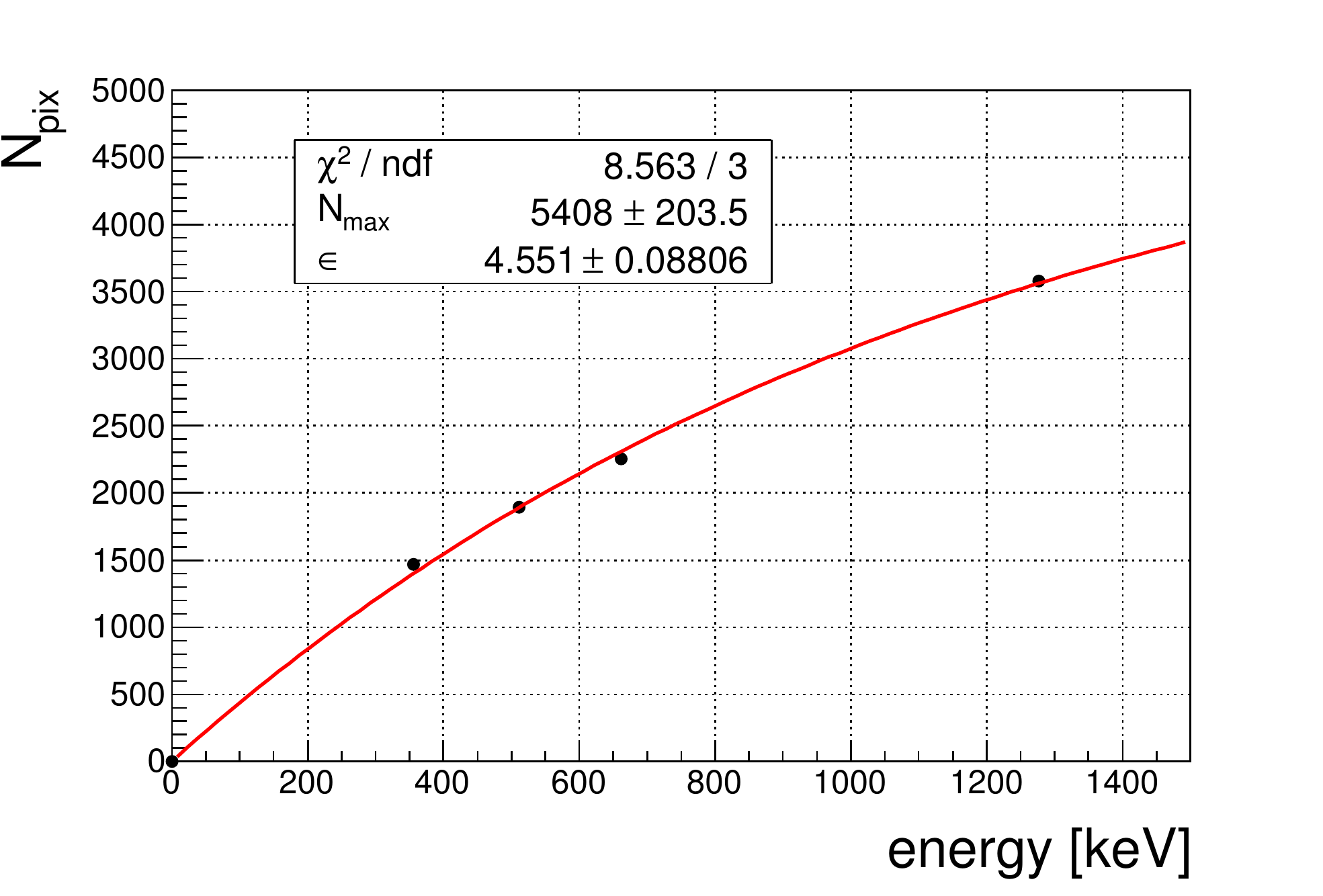}
\caption{Detector light output for different gamma energies, the data
  are fitted with equation \ref{eq:sat}.}
\label{fig:EcalFit}
\end{center}
\end{figure}

\begin{figure}[H]
\begin{center}
\includegraphics[width=\PicSize\columnwidth]{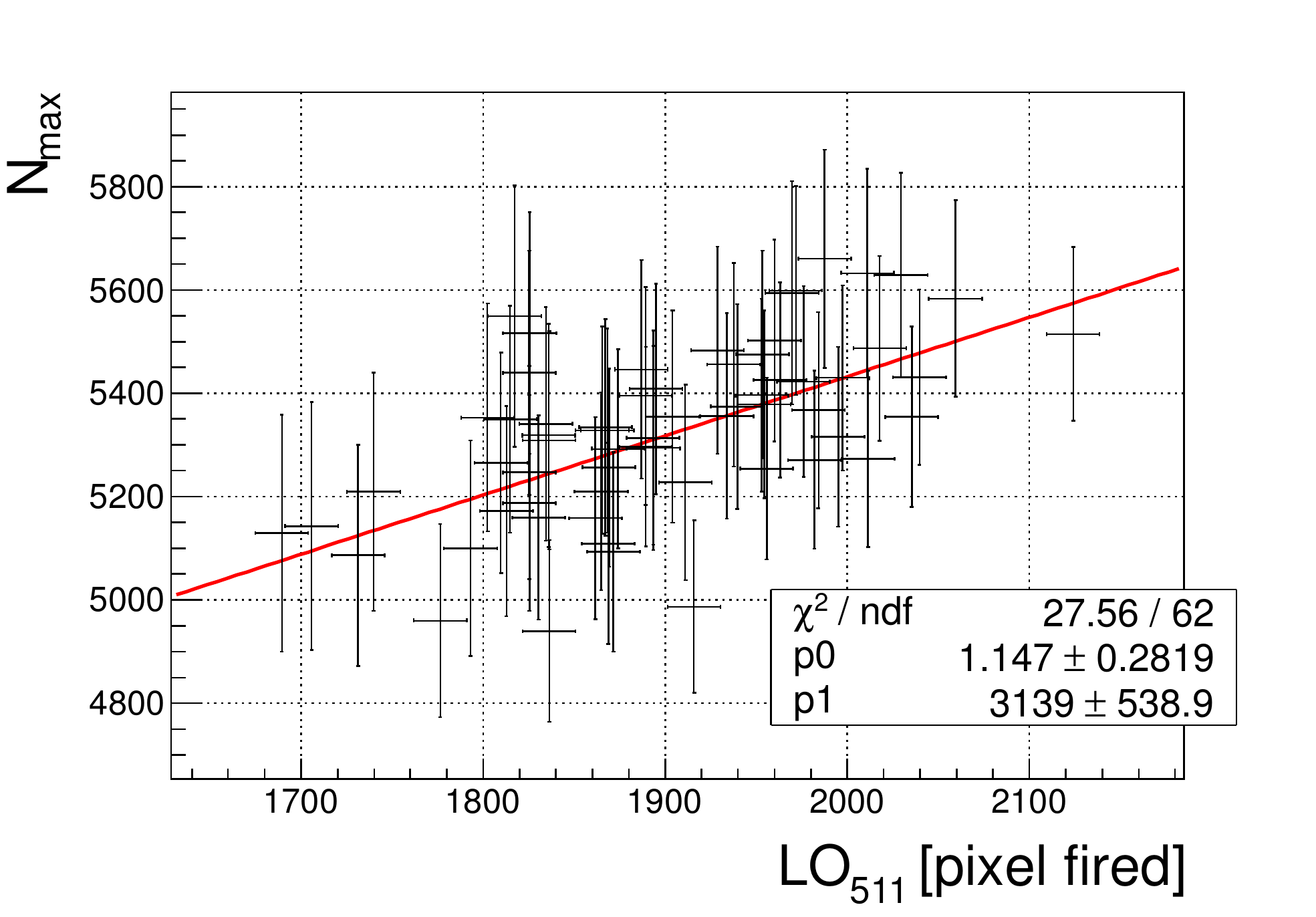}
\caption{Distribution of the parameter $N_{max}$ of equation \ref{eq:sat}
  depending on $LO_{511}$ for 64 channels. The red line is the linear fit.}
\label{fig:P0}
\end{center}
\end{figure}

\begin{figure}[H]
\begin{center}
\includegraphics[width=\PicSize\columnwidth]{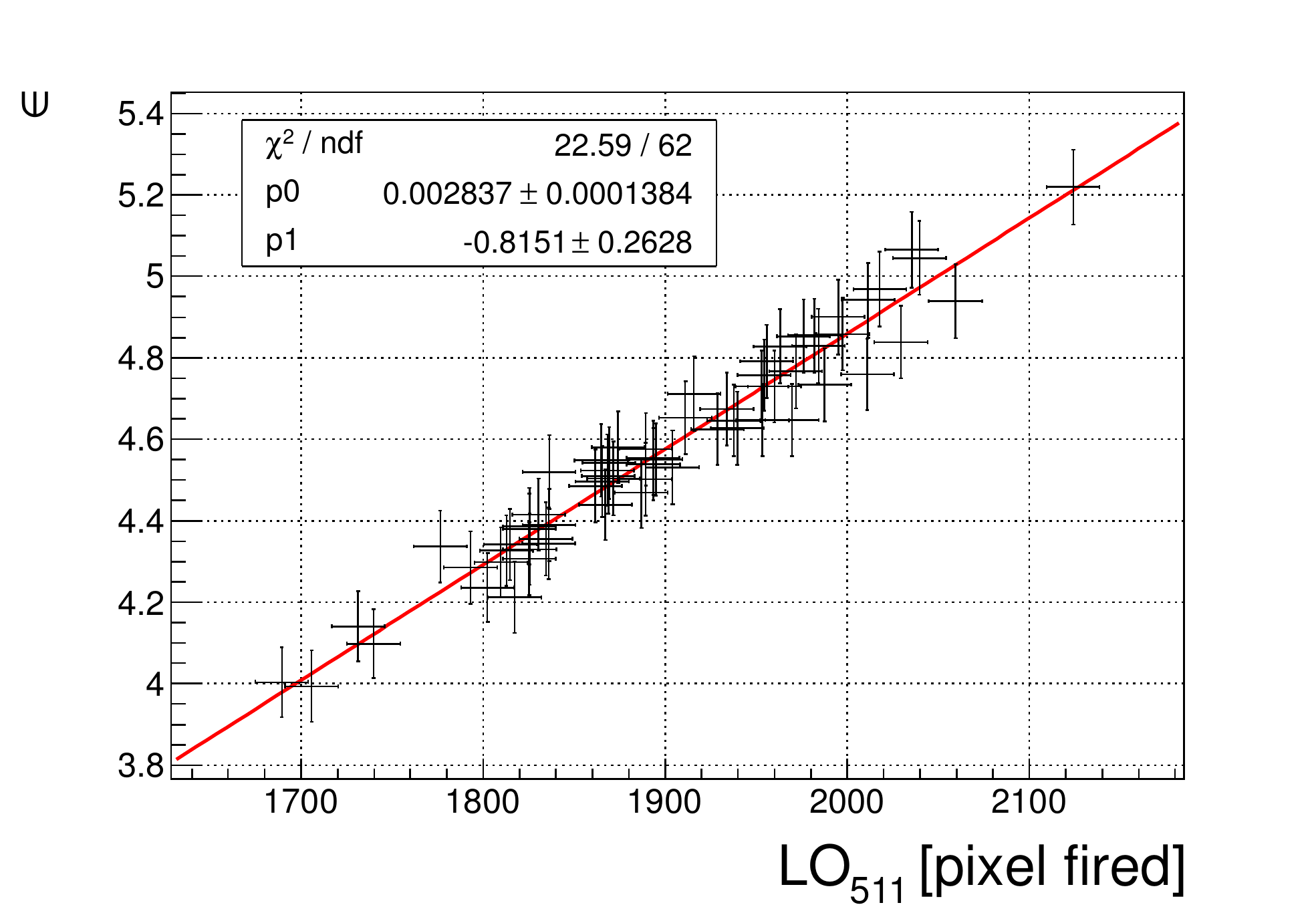}
\caption{Distribution of the parameter $\epsilon$ of equation \ref{eq:sat}
  depending on $LO_{511}$ for 64 channels. The red line is the linear fit.}
\label{fig:P1}
\end{center}
\end{figure}

\begin{figure}[H]
\begin{center}
\includegraphics[width=\PicSize\columnwidth]{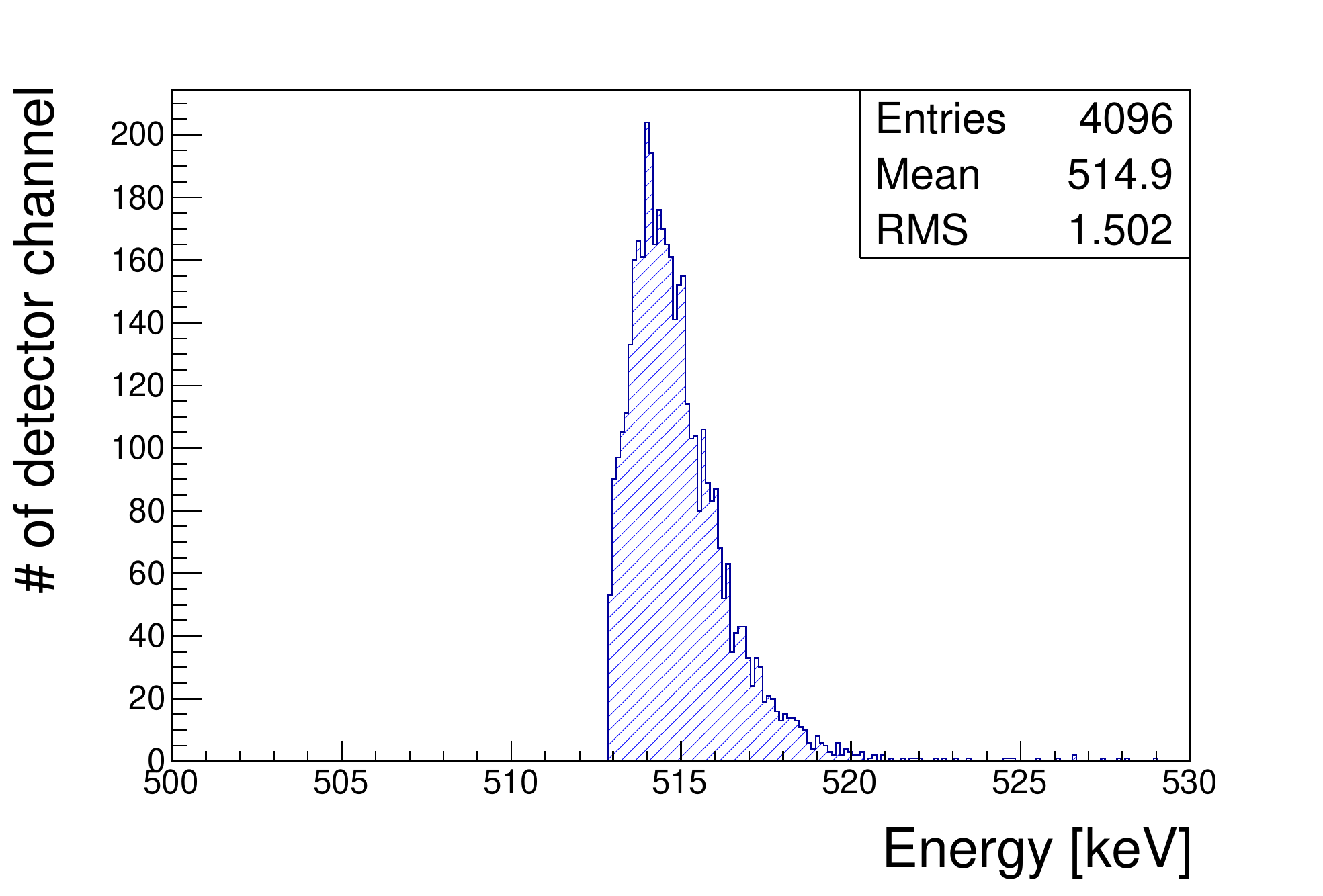}
\caption{Energy Distribution calculated from the position of the 511 keV peak (from $^{22}$Na) for all the channels after
  the energy correction has been applied. The tail on the right
  correspond to the channels with very low $LO_{511}$.}
\label{fig:E511}
\end{center}
\end{figure}

\begin{figure}[H]
\begin{center}
\includegraphics[width=\PicSize\columnwidth]{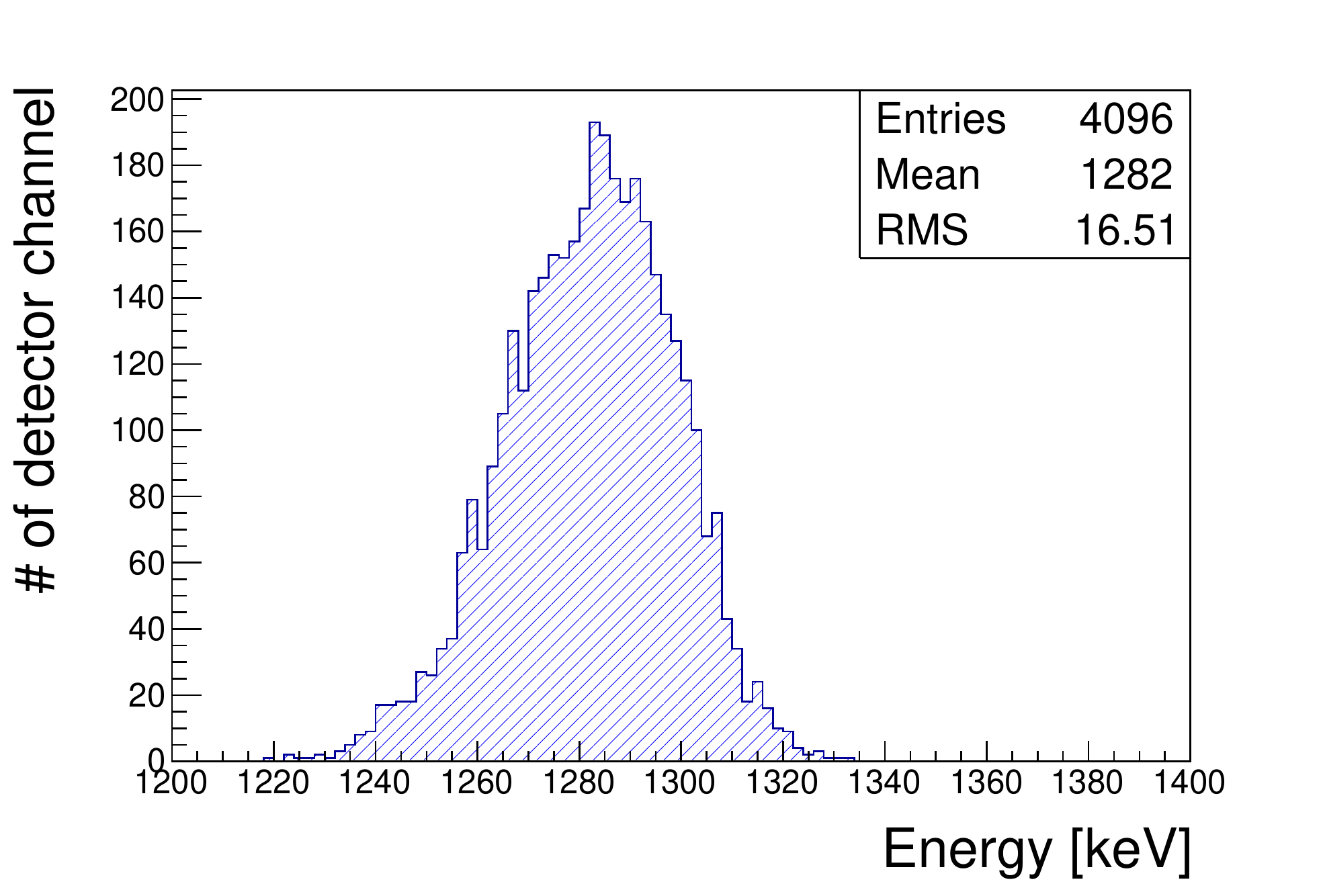}
\caption{Energy Distribution calculated from the position of the 1277 keV peak (from $^{22}$Na) for all the channels after
  the energy correction has been applied.}
\label{fig:E1277}
\end{center}
\end{figure}

\begin{figure}[H]
\begin{center}
\includegraphics[width=\PicSize\columnwidth]{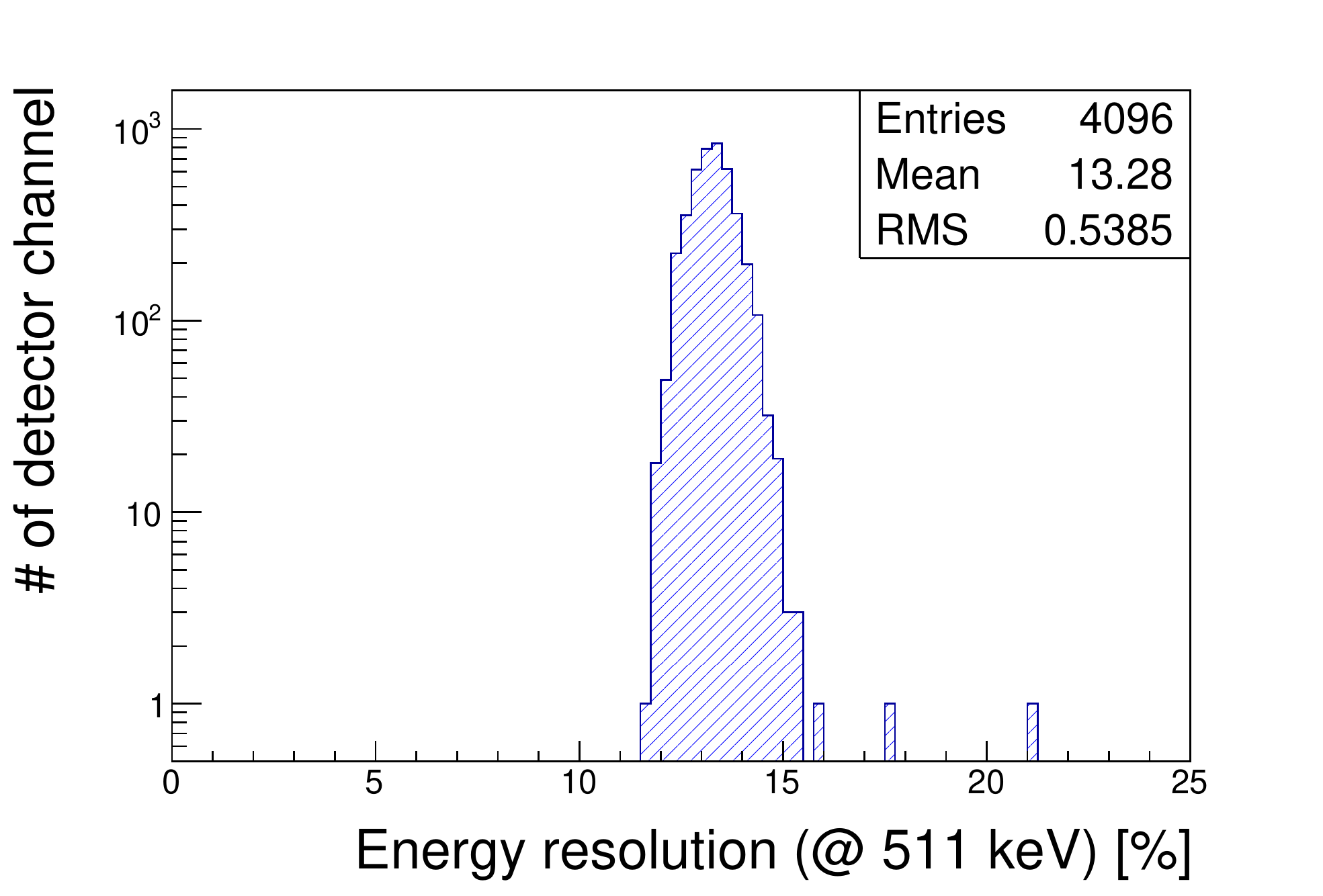}
\caption{Energy resolution at 511 keV for all the channels.}
\label{fig:Eresol}
\end{center}
\end{figure}

It is well known that the SiPM is a non-linear device due to the
limited number of pixels, therefore a non-linear correction is
necessary in order to estimate the energy resolution for the 511 keV
gamma. \\
Using the setup described in section \ref{sec:LYmodule}, the detector light output is measured for different
gamma energies:  356 keV from $^{133}$Ba, 511 and 1277 keV from $^{22}$Na, and
662 keV from $^{137}$Cs. \\
The measurement is performed on a sample of 4 modules (64 channels):
for each channel the light output data are fitted with the
function describing the SiPM saturation:
\begin{equation}
N_{pix}=N_{max}\left(1-e^{-\frac{\epsilon \cdot E}{N_{max}}}\right),
\label{eq:sat}
\end{equation}
where $N_{pix}$ is the average number of pixels fired on the photodetector,
$\epsilon$ is the number of pixel fired on the SiPM per unit of
energy deposited in the crystal and $E$ is the energy of the
detected gamma photon. $N_{max}$ is related to the maximum number of pixels in the SiPM. This
number is actually higher than the nominal value (3464 pixels) because
the pixel recovery time is about 20 ns, which is half of the decay
constant of the crystal (40 ns). \\
Figure \ref{fig:EcalFit} shows the non-linear function for one channel,
however this function is not unique for all the channels.
It is expected that the parameter $\epsilon$ varies among the channels
according to $LO_{511}$, which takes into account the variations in
the SiPM properties (photo detection efficiency, correlated noise),
crystal light yield and the optical coupling. \\
The dependence of $N_{max}$ and $\epsilon$ to the $LO_{511}$ for the 64
channels is shown in Figure \ref{fig:P0} and \ref{fig:P1} respectively. Although the linear
dependence of $\epsilon$ was expected, also $N_{max}$ shows a dependence on
$LO_{511}$, maybe due to variations in in the pixel recovery time or
crystal decay time.
In both cases a linear fit is applied and the relations obtained are
used to calculate $N_{max}$ and $\epsilon$ for any other channel depending on its
$LO_{511}$. \\
Figure \ref{fig:E511} shows the energy obtained using equation
\ref{eq:sat} and the $LO_{511}$ for every channel. Similarly, the corresponding
energy of the 1277 keV peak of the $^{22}$Na spectrum is estimated
for every channel (Figure \ref{fig:E1277}). In both cases the predicted
energy is in good agreement with the measurements, the precision
worsens only if the channel $LO_{511}$ is significantly lower than the
average. \\
Finally, after the energy calibration has been applied, the energy
resolution for the 511 keV gamma is evaluated for all the
channels (Figure \ref{fig:Eresol}). Only one channel is out of the detector requirement: at least 20\%
is necessary to effectively discriminate the 511 keV gammas from the
Compton events.

\subsection{Coincidence Time Resolution}
\begin{figure}[H]
\begin{center}
\includegraphics[width=\PicSize\columnwidth]{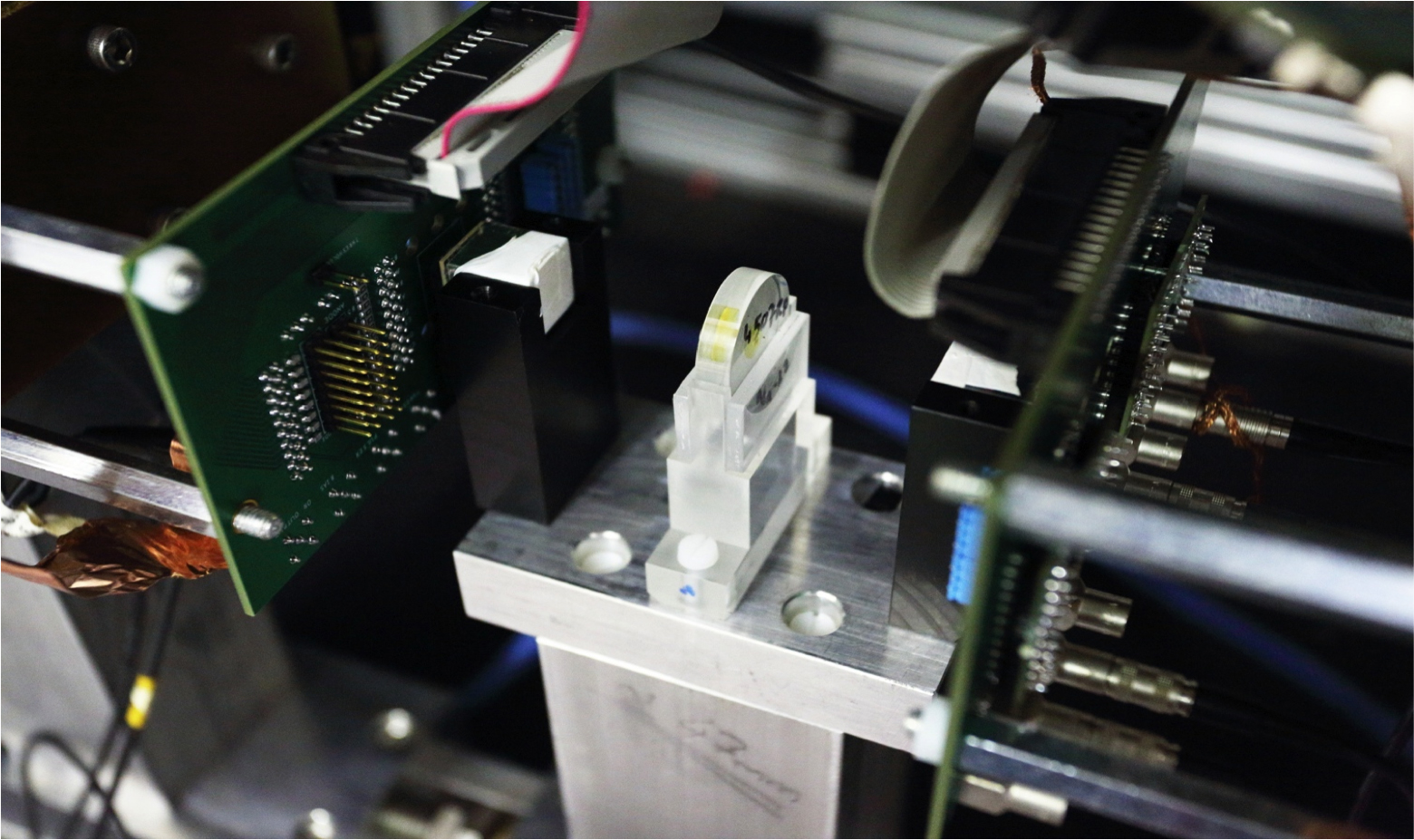}
\caption{The setup used for the timing measurements. The two
detector modules are placed facing each other, 10 cm apart. A
$^{22}$Na source is placed between the two modules.}
\label{fig:ctr_setup}
\end{center}
\end{figure}

\begin{figure}[H]
\begin{center}
\includegraphics[width=1.01\columnwidth]{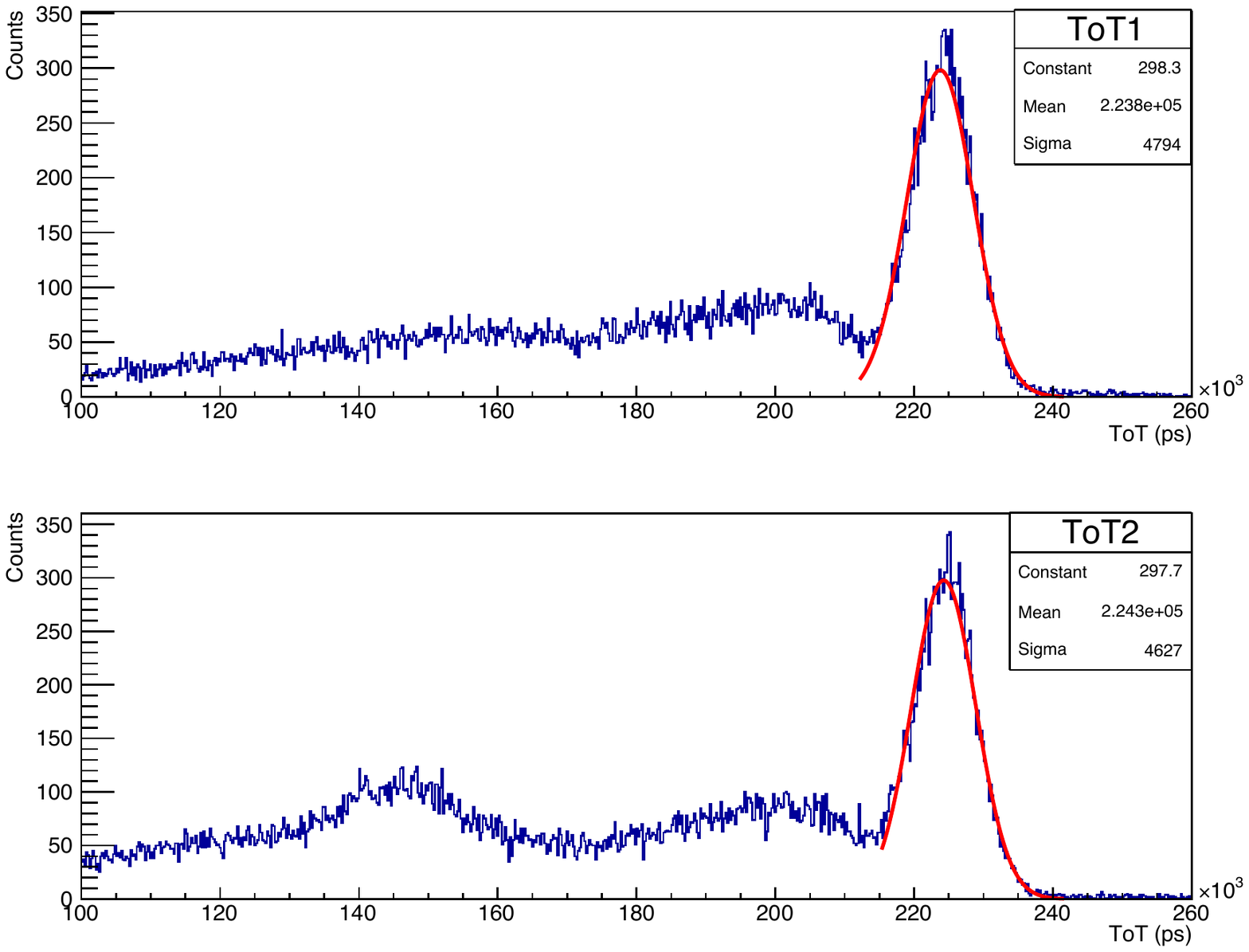}
\caption{Typical Time over Threshold spectra for two channels in coincidence.}
\label{fig:ctr_TOTspectrum}
\end{center}
\end{figure}

\begin{figure}[H]
\begin{center}
\includegraphics[width=0.68\columnwidth]{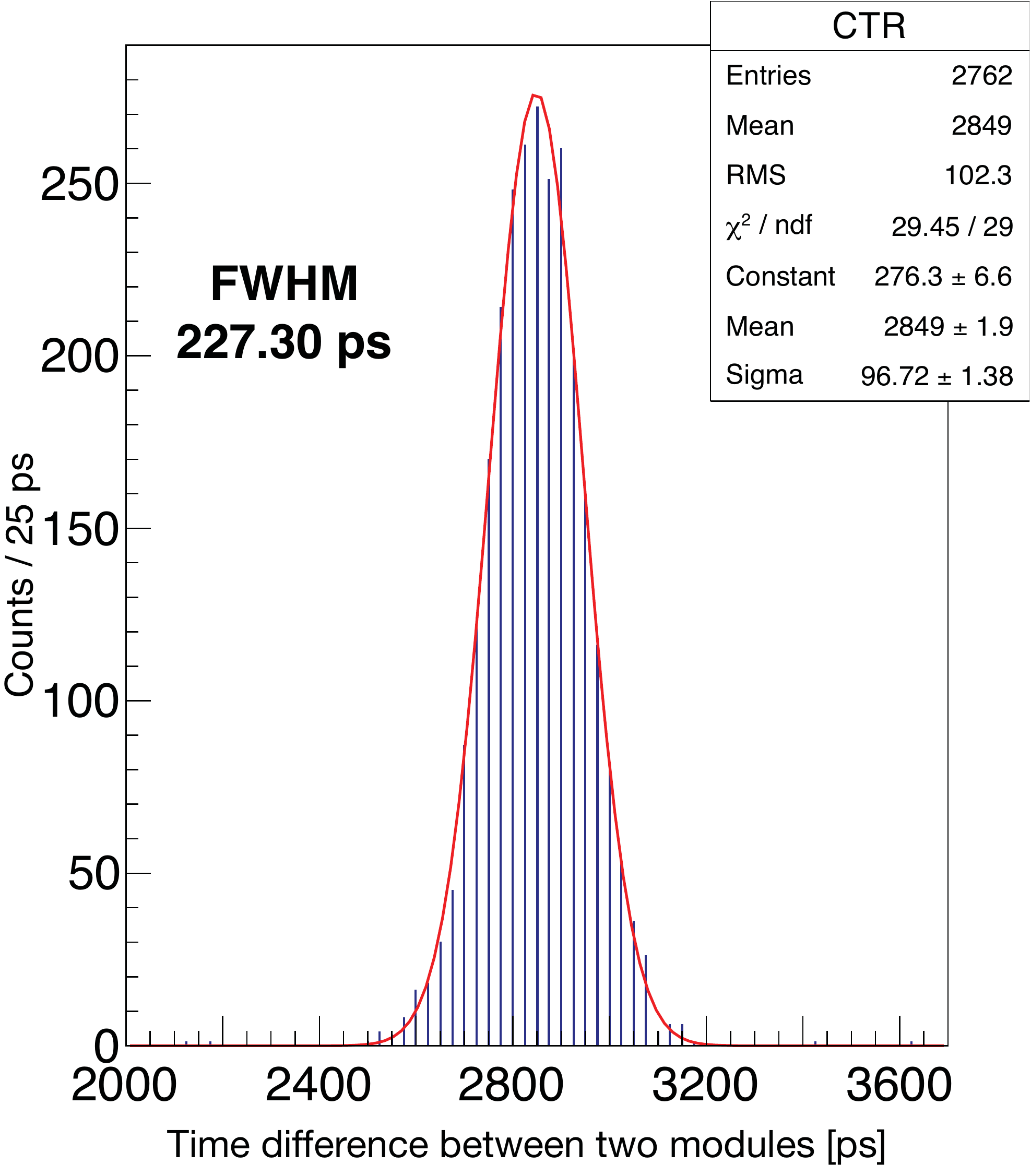}
\caption{The CTR is the FWHM of a Gaussian fit of the time difference
between each channel pair.}
\label{fig:ctr_Tspectrum}
\end{center}
\end{figure}

\begin{figure}[H]
\begin{center}
\includegraphics[width=\PicSize\columnwidth]{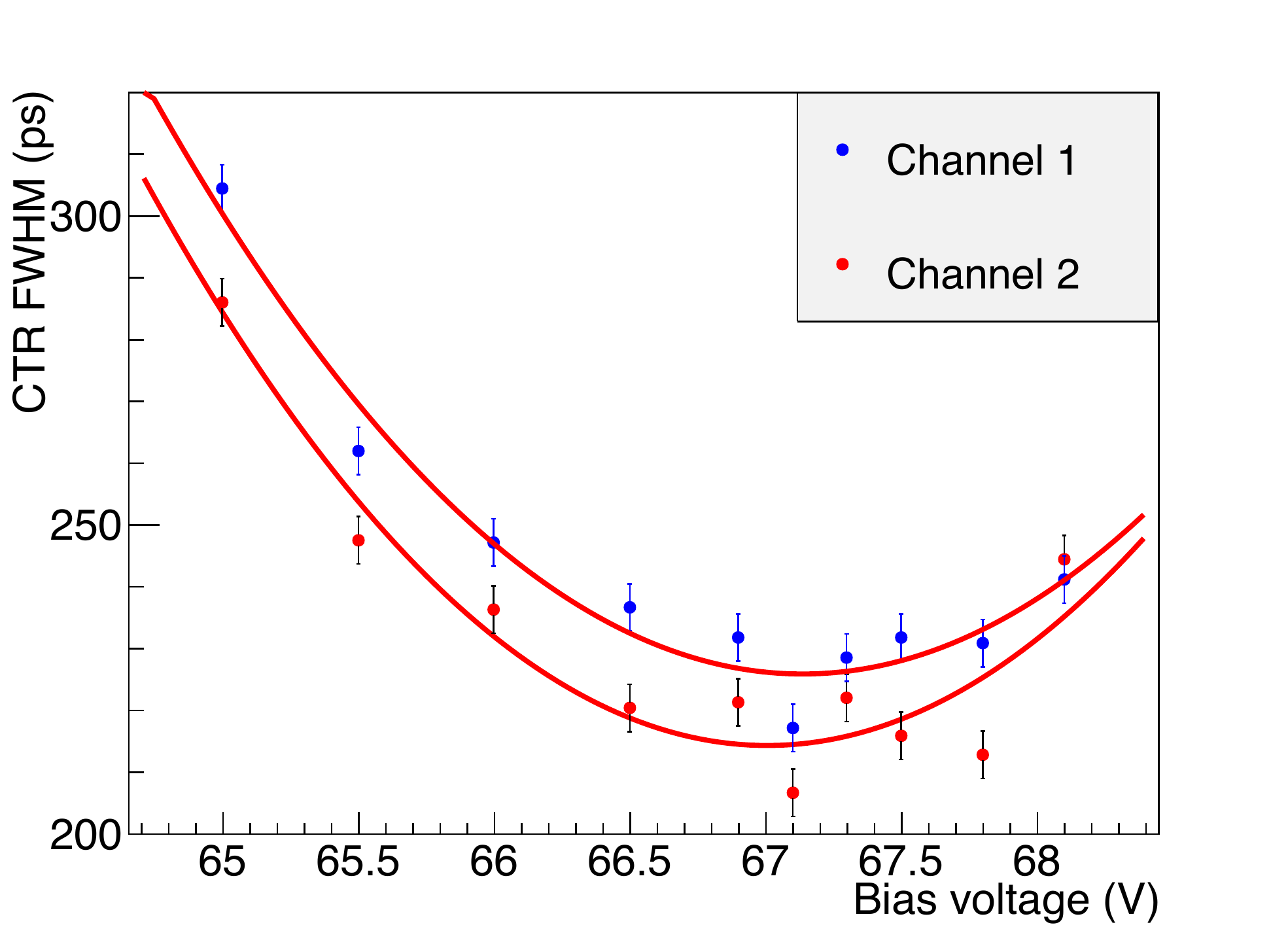}
\caption{CTR as function of bias voltage for one pair of
  channels with respect to the reference module. Temperature fixed at 19 $^o$C.}
\label{fig:ctr_Vscan}
\end{center}
\end{figure}

\begin{figure}[H]
\begin{center}
\includegraphics[width=\PicSize\columnwidth]{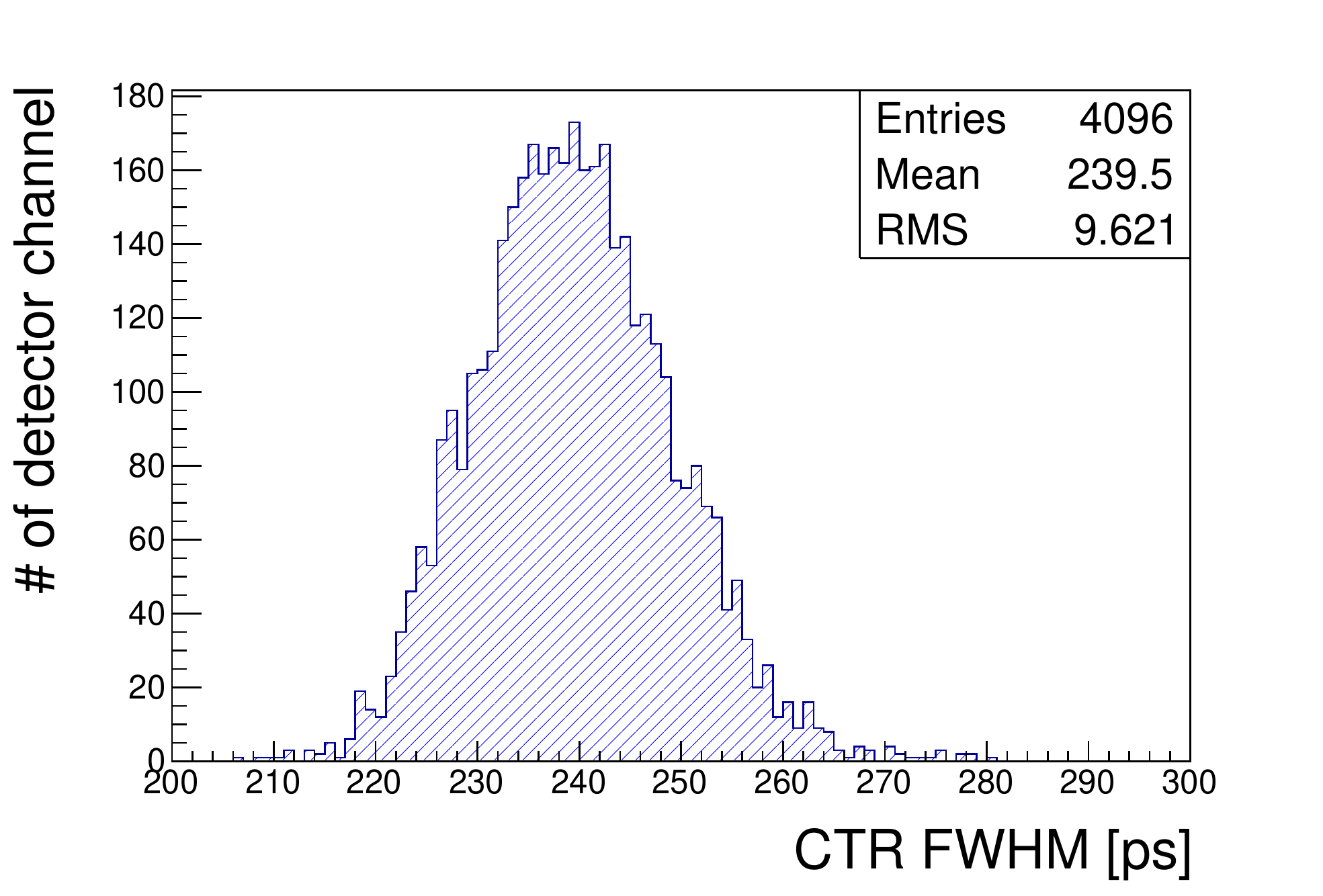}
\caption{The CTR distribution of all channels pairs, at 19 $^o$C and at 2.5 V excess bias.}
\label{fig:ctr_all}
\end{center}
\end{figure}

The coincidence time resolution of all the channels is measured using the setup shown in Figure \ref{fig:ctr_setup}.
The two modules are facing each other at 10 cm distance, while a $^{22}$Na point
like source is placed between them on the line connecting the module centers. One module has been selected as the reference for the measurement
of the other 256 modules. \\
The output signal of the SiPM is amplified and discriminated by NINO ASIC \cite{nino} and
sent to a high precision TDC (HPTDC- 25ps LSB) \cite{TDC}.
The data analysis is performed with a total number of 4x10$^6$
triggers. As shown in Figure \ref{fig:ctr_TOTspectrum}, the
Time-over-Threshold (ToT) spectra are obtained for any two channels in
coincidence; then a gaussian fit is applied to the peaks corresponding to
the 511 keV gamma events.
Hence the time difference for the gamma events of the
two channels are histogrammed and shown in Figure \ref{fig:ctr_Tspectrum}. A Gaussian
fit is applied and the FWHM is taken as the CTR of the channel
measured (227.3 ps $\pm$ 3.2 ps in this case). \\
Figure \ref{fig:ctr_Vscan} shows the CTR as a function of bias voltage
for two channels in coincidence with the reference module. Due to the increase of the SiPM Photon Detection
Efficiency (PDE) with the increase of excess bias, it is expected that
also the CTR improves accordingly. However, also the DCR increase with
the excess bias, and at a given point it becomes predominant. 
In order to compare the CTR of all the 256 modules, an excess bias of 2.5 V is used
for any channel, the temperature is fixed at 19 $^o$C and the NINO
ASIC threshold is kept fixed at
55 mV, which is equivalent to the threshold of 0.5 photoelectron. 
The CTR distribution for all the channels, measured with respect to
the reference module, is presented in
Figure \ref{fig:ctr_all} and shows an average of 239 $\pm$ 10 ps FWHM.

\section{Conclusions}
In the frame of the EndoTOFPET-US project, all the detector modules
for the external plate have been characterized.
The gain, breakdown voltage, DCR and correlated noise have been
measured for all the received SiPMs, only 2\% of them have been
rejected due to excess DCR.\\
The LYSO:Ce crystal matrices show a very good light yield and a
promising average CTR of 239 $\pm$ 10 ps FWHM has been obtained, which
can be further improved in ideal conditions, such as lower
temperature or optimal excess bias for each channel. The average energy
resolution (at 511 keV) for all the modules is about 13\%, and it complies with the minimum requirement of 20\%. The light output
for every channel is available for a preliminary detector calibration, and
an average of about 1800 pixels fired for a 511 keV gamma
interaction has been obtained. \\
The detector modules are now ready for the integration with the
dedicated ASICs and the final mechanical assembly.


\section*{Acknowledgements}
The research leading to these results has received funding from the European Union Seventh
Framework Programme [FP7/2007-2013] under Grant Agreement no. 256984, and is supported
by a Marie Curie Early Initial Training Network Fellowship of the European Community's Seventh Framework Programme under contract number (PITN-GA-2011-289355-PicoSEC-MCNet).



\end{multicols}

\end{document}